\documentclass[prx,twocolumn,english,superscriptaddress,longbibliography]{revtex4-2}

\usepackage{enumerate}
\usepackage{amsfonts,amssymb,amsmath,physics}
\usepackage[]{graphics,graphicx,epsfig}
\usepackage{amsthm}
\usepackage{graphicx}
\usepackage{dcolumn}
\usepackage{natbib}
\usepackage{color}
\usepackage{multirow}
\usepackage{ulem}
\usepackage{bm}
\usepackage{url}
\usepackage{braket}
\usepackage[caption=false]{subfig}

\usepackage{siunitx}
\usepackage{mathtools,leftidx}
\usepackage{dsfont}
\usepackage{chngcntr}

\usepackage{tikz}
\usetikzlibrary{quantikz}

\usepackage{hyperref}
\usepackage[noabbrev]{cleveref}

\begin{document}

\title{Cubic Phase Gate with a Trapped Ion Oscillator}

\author{Nigel Benjamin Lee Junsheng}
\affiliation{Centre for Quantum Technologies,
National University of Singapore, 3 Science Drive 2, 117543 Singapore,
Singapore}

\author{Eugene Koh Jun Wei}
\affiliation{Centre for Quantum Technologies,
National University of Singapore, 3 Science Drive 2, 117543 Singapore,
Singapore}

\author{Kim Mu Young}
\affiliation{Centre for Quantum Technologies,
National University of Singapore, 3 Science Drive 2, 117543 Singapore,
Singapore}

\author{Dzmitry Matsukevich}
\affiliation{Centre for Quantum Technologies,
National University of Singapore, 3 Science Drive 2, 117543 Singapore,
Singapore}
\affiliation{Department of Physics,
National University of Singapore, 3 Science Drive 2, 117543 Singapore,
Singapore}

\date{\today}

\begin{abstract}
The motional modes of trapped ions are promising candidates for continuous-variable (CV) quantum computation and simulation. However, to achieve a universal gate set for CV quantum computation, anharmonic nonlinear unitaries are required. Here, we experimentally demonstrate a protocol that allows us to approximate the cubic phase gate. By numerically optimizing the phase angles and amplitudes of a set of state-dependent force pulses, we demonstrate the cubic phase gate for both vacuum and low-$\alpha$ coherent state inputs and verify its performance. Our protocol can also be extended to approximate other nonlinear unitaries. 

\end{abstract}

\maketitle

Motional modes of trapped ions are a promising candidate for the realization of continuous-variable (CV) quantum computing \cite{ion1,ion2,ion3,ion4,ion5,ion6,ion7,ion8,ion9,GKP,ion11}. CV quantum computation and simulation utilize the infinite-dimensional Hilbert space of bosonic modes \cite{application1,application2,application3}, which could significantly reduce hardware requirements \cite{overhead1,overhead2} compared to the traditional discrete variable approach \cite{nielsen_quantum_2010}. 

In addition to single-mode linear bosonic gates (displacement, phase rotation, and squeezing) and a two-mode bosonic gate (beamsplitter), nonlinear gates are necessary for realizing a universal set of quantum gates for a continuous-variable quantum computer \cite{braunstein1999, Gottesman2001}. However, efficiently implementing such nonlinear, anharmonic couplings without resorting to tailored hardware \cite{hardware1,hardware2,hardware3,hardware4,harlander_trapped-ion_2011} remains a challenge. 

One example of a nonlinear gate is the cubic phase gate (CPG), which is described by the unitary $U_{CPG}=\exp(i\kappa\hat{X}^3)$~\cite{PhysRevResearch.6.023332,PhysRevA.79.062318}. Although several methods to implement such a gate in various physical systems have been  proposed~\cite{PhysRevA.91.032321,PhysRevA.95.052352,PhysRevA.93.022301,QubicPhaseGate_QML_proposal,PhysRevLett.125.160501}, experimental demonstration remains a difficult task for most platforms~\cite{superconducting_CPG, fontaine2026programminganharmonicpotentialssuperconducting}. 

A recent proposal by K. Park and R. Filip~\cite{filip2024} suggested that the cubic phase gate can be approximated via a finite Fourier series expansion. It takes advantage of the nonlinear coupling between the CV oscillator and the discrete-variable (DV) two-level qubit system and a series of spin-dependent displacement pulses to emulate the Fourier coefficients. This proposal is similar to bosonic quantum signal processing (QSP), which is a programmable protocol that allows for the generation of arbitrary nonlinear unitaries \cite{weibe2024}. Since bosonic QSP schemes require hybrid interactions between CV-DV systems, trapped-ion quantum processors are a feasible platform to realize these schemes. Bosonic QSP schemes have garnered significant interest recently and have found applications in quantum simulation \cite{tingrei2025,tingrei2026,QSP_uracil,fontaine2026programminganharmonicpotentialssuperconducting}, continuous-variable quantum computing \cite{filip2024,GKP,QFT,fong2025engineeringnongaussianbosonicgates}, quantum metrology \cite{QSP_sensing1,QSP_sensing2,QSP_sensing3}, and device control/characterization \cite{QSP_device1,majumdar2026analyticapproachquantumcontrol}.

In this paper, we demonstrate an efficient, easily reproducible protocol to approximate cubic nonlinearity on the motional modes of our trapped-ion platform. The length of the protocol is short compared to the coherence time of the bosonic mode. Our protocol can also be customized to generate other nonlinear gates.

\section{THEORETICAL FRAMEWORK} \label{THEORETICAL FRAMEWORK}

\begin{figure*}[t]
    \centering
    \resizebox{0.99\textwidth}{!}{
    \begin{quantikz}[thin lines,node distance=2pt] 
            \lstick{$\ket{g}$}& \qw  &  \gate{R_{y}^\dagger(\pi/2)} \gategroup[wires=2,steps=7,style={dotted,cap=round,inner sep=10pt}]{Cubic Phase Gate Protocol} &\ctrl{1} &\ctrl{1} & \qw &  \ldots\  &\ctrl{1}   &  \gate{R_{y}(\pi/2)} &\qw &  \gate{R_x(\pi/2)} \gategroup[wires=2,steps=3,style={dotted,cap=round,inner sep=10pt}]{Characteristic Function Measurement} & \ctrl{1} &\meter{} \\
            \lstick{$\ket{0}$} & \qw &\qw    & \gate{SDF(\phi_s=\phi_0,\phi_m=0,\xi_0)}  & \gate{SDF(\phi_1,0,\xi_1)} & \qw &  \ldots\ & \gate{SDF(\phi_{N-1},0,\xi_{N-1})} & \qw &\qw &\qw &  \gate{SDF(\xi_{\chi})} &\qw
    \end{quantikz}
    }
    \caption{Illustrated circuit diagram of the experimental protocol. It requires the access to an ancillary qubit in addition to the chosen bosonic mode which we assume starts off in the ground state $\ket{g}$ and vacuum state $\ket{n=0}$ respectively. It consists of a single qubit rotation ($R_{y}^\dagger$) gate to bring the qubit spin state to $\ket{-}$, followed by $N$ SDF pulses onto the bosonic mode. The spin phase ($\phi_i$) and displacement amplitude ($\xi_i$) of each SDF pulse is set to the values that have been previously found via numerical optimization while the motional phase ($\phi_m$) values are set to zero. Another $R_{y}$ gate is implemented to bring the qubit back to the computational basis. At the end, we implement a characteristic function measurement circuit as per \cite{home2020} which consists of a single qubit rotation gate ($R_x$) and another SDF pulse (more details available in the Supplementary Material).}
    \label{fig:circuit}
\end{figure*}
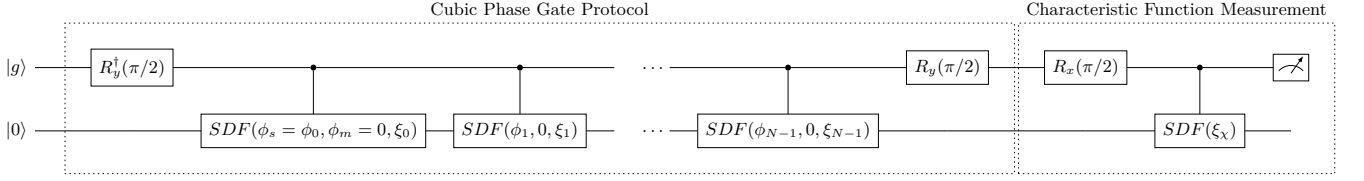

We first simulate the unitary evolution of a vacuum state under the cubic phase gate, which would generate the cubic phase state described by $\ket{\kappa}=U_{CPG}\ket{0}=\exp(i\kappa\hat{X}^3)\ket{0}$, where $\ket{0}$ denotes the vacuum Fock state and $\kappa$ denotes the parameter of the cubic phase gate. Here, we define the arbitrary quadrature $\hat{Q}(\theta)=(\hat{a} e^{-i\theta}+\hat{a}^\dagger e^{i\theta})/\sqrt{2}$ where $\hat{a}$($\hat{a}^\dagger$) is the annihilation(creation) operator with $\theta\in[0,2\pi)$. We also define the position $(\hat{X}=\hat{Q}(0))$ and momentum $(\hat{P}=\hat{Q}(\frac{\pi}{2}))$ operators.

{\color{black}
To approximate a cubic phase gate, we apply a sequence of $N$ spin-dependent-force (SDF) pulses,
\begin{align}
\label{product_operator}
\hat U_{\mathrm{CPG}}
\approx \hat U
&=
\prod_{j=1}^{N}
\mathrm{S\hat{D}F}(\phi_j,\xi_j),
\end{align}
Each pulse is described by the unitary operator

\begin{equation}
\label{eq:SDF}
\mathrm{S\hat{D}F}(\phi,\xi) = \exp\left( i\sqrt{2}\xi \hat{\sigma}_{\phi}\hat X \right),
\end{equation}
where
\begin{equation}
\hat{\sigma}_{\phi} \equiv \cos(\phi)\hat{\sigma}_x + \sin(\phi)\hat{\sigma}_y, \qquad \hat X = \frac{\hat a+\hat a^\dagger}{\sqrt{2}}.
\end{equation}

In general, the total unitary can be written as
\begin{equation}\label{total_u}
\hat U = A(\hat X)\hat I -i\left[ B_x(\hat X)\hat{\sigma}_x + B_y(\hat X)\hat{\sigma}_y + B_z(\hat X)\hat{\sigma}_z \right].
\end{equation}
Imposing the additional requirement that a qubit initially prepared in the $\ket{+}$ eigenstate of the $\hat{\sigma}_x$ operator returns to the same state at the end of the sequence would ensure that $B_y(\hat X)=B_z(\hat X)=0$ and that the total unitary takes on the form:
\begin{equation}
\hat U
=
\exp\left[-i\Phi(\hat X)\hat{\sigma}_x\right].
\end{equation}
If the qubit was initially in $|-\rangle$, the oscillator instead undergoes unitary operation $\exp[i\Phi(\hat X)]$. Here, $\Phi(\hat X)$ is a real odd function of $\hat X$. A cubic phase gate is obtained when $\Phi(\hat X)$ approximates a function proportional to $\hat X^3$. 
}

We experimentally implement the SDF with a bichromatic laser field that realizes the interaction Hamiltonian \cite{monroe2018}: 

\begin{multline} \label{eq:H_I}
    H_I(t)
    = -\frac{\hbar}{2}\Omega(e^{-i\phi_s}\hat{\sigma}_{+}+e^{i\phi_s}\hat{\sigma}_-) \\\times(\hat{a}e^{i\delta t+i\phi_m} + \hat{a}^\dagger e^{-i\delta t-i\phi_m}),
\end{multline}

\noindent where $\phi_s$ and $\phi_m$ are known as the spin phase and motion phase terms, $\Omega_0$ is the Rabi frequency, $\hat{\sigma}_\pm=(\hat{\sigma}_x\pm i\hat{\sigma}_y)/2$ are the qubit raising(+)/lowering(-) operators and $\delta$ is the laser frequency detuning away from the carrier frequency $\nu_c$. More details available in Appendix \ref{appendix:appendix1} and the Supplementary Material.

\begin{figure*}[t]
\includegraphics[width=0.99\textwidth]{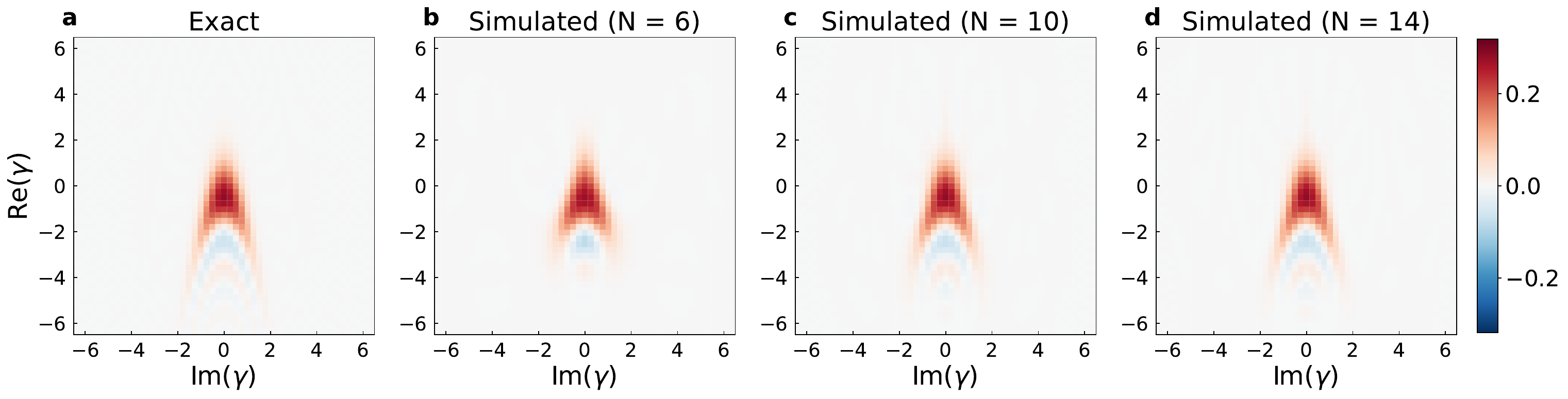}
\caption{Wigner functions of an initial vacuum state after (a) evolving under the cubic unitary where $\kappa=-1/2$ and (b-d) after evolving under our protocol with our numerically optimized parameters for different values of $N$. Output from our protocol where (b) $N=6$ (c) $N=10$ (d) $N=14$ results in a simulated state overlap of (b) $0.983$ (c) $0.995$ (d) $0.998$ when compared with (a). } 
\label{fig:theory_allSim}
\end{figure*}

\begin{figure}[t]
\includegraphics[width=0.45\textwidth]{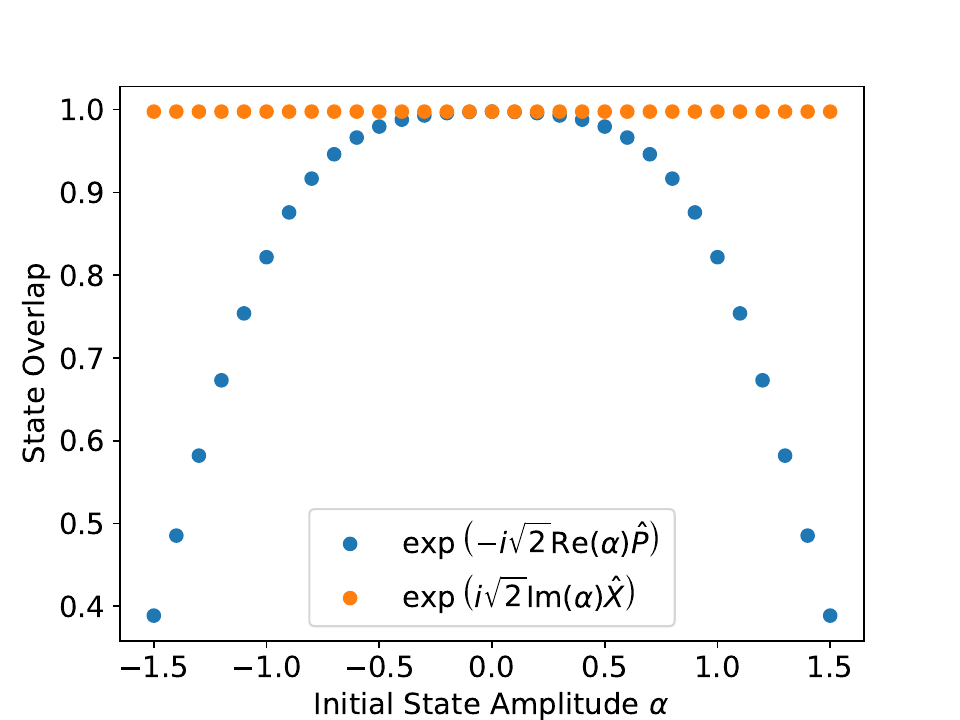}
\caption{Plot of state overlap between the cubic unitary evolution and the $N=14$ protocol evolution for different initial coherent states. The overlaps were calculated by subjecting an initial vacuum state to coherent displacements generated by $D_{P}(\alpha)=\exp\left(-i\sqrt{2}\operatorname{Re}(\alpha)\hat{P}\right)$ (blue markers) or generated by $D_{X}(\alpha)=\exp\left(i\sqrt{2}\operatorname{Im}(\alpha)\hat{X}\right)$ (orange markers) for various displacement amplitudes. Following which, the prepared state is either evolved under the cubic unitary or the $N=14$ protocol. Finally, the state overlap between these two cases were calculated. } 
\label{fig:initial_state_fidelity}
\end{figure}

\section{Numerical Optimization} \label{sec:Numerical Optimization}

Following the proposal \cite{filip2024}, we approximate evolution under a cubic potential using a sequence of SDF pulses, each parameterized by its gate angle $\phi_s\equiv\phi_i$ and amplitude $\xi_i$. The index $i$ labels the current pulse number and runs from $0$ to $N-1$ where $N$ is the length of the sequence. To find the optimal set of parameters, we first initialize our bosonic mode to the vacuum state and the qubit in the $\sigma_x$ eigenstate $\ket{-}=(\ket{g}-\ket{e})/\sqrt{2}$, where $\ket{g}$ and $\ket{e}$ refer to the spin-down and spin-up qubit states respectively. We then minimize $1-|\braket{\psi_{tar}|\psi_{sim}}|$, where $\ket{\psi_{\mathrm{tar}}}$ is the target motional state and $\ket{\psi_{\mathrm{sim}}}$ is the motional state produced by the simulated protocol. The initial guesses for the parameters were randomly generated between $0$ to $2\pi$ for $\phi_i$ and between $0$ and $1$ for $\xi_i$. After the sequence, an additional rotation pulse returns the qubit to the computational basis. The complete protocol is illustrated in Figure \ref{fig:circuit}. 

To reduce the number of optimization parameters from $2N$ to $N$, we impose symmetry constraints on $\phi_i$ and $\xi_i$ during numerical optimization. For example, for a protocol of length $N=10$, the angles follow the pattern, $\phi_3=\phi_5$, $\phi_2=\phi_6$ and so on until $\phi_0=\phi_8$. The angle for $\phi_4$ will have a unique value and $\phi_9=0$. The values of $\xi_i$ are also symmetric in a slightly different way, where $\xi_4=\xi_5$, $\xi_3=\xi_6$, and so on until $\xi_0=\xi_9$. The optimization procedure ensures that the contribution of $B_y(X)$ and $B_z(X)$ to the final state is small compared to $B_x(X)$ because the target state of the harmonic oscillator, after tracing out the qubit, is a pure state. At the end of the sequence of SDF pulses, the qubit is almost perfectly unentangled from the harmonic oscillator and returns to a near-pure (purity of $0.9992$ for $N=14$) qubit state which is close to $\ket{-}$ ($\{\sigma_x,\sigma_y,\sigma_z\}=\{-0.9992,-0.0014,0\}$ for $N=14$). The optimized parameters used in this work and the final qubit states are provided in the Supplementary Material.

Figure \ref{fig:theory_allSim} shows the simulated Wigner functions for the optimized protocols with $N=6,10,$ and $14$ pulses. We compare each of these three Wigner functions with the Wigner function generated via cubic unitary time evolution (with $\kappa=-1/2$) and calculate the corresponding state overlaps. As expected, the state overlaps improve with increasing $N$. 

To demonstrate that our protocol has successfully approximated the CPG, rather than merely producing the cubic phase state from vacuum, we simulate the same $N=14$ protocol using coherent state inputs instead. Figure \ref{fig:initial_state_fidelity} shows the state overlap for various values of coherent state amplitude $\alpha_{initial}$. For $|\alpha_{initial}| \lesssim 0.5$, the overlap is lower than vacuum but still above $0.95$. However, for $|\alpha_{initial}| \gtrsim 0.5$ along the $\hat{P}$ axis, the state overlaps decay rapidly. Since displacements along the $\hat{X}$ axis commutes with SDFs in the CPG protocol, the state overlaps do not change over any displacement along the $\hat{X}$ axis. We experimentally test four input states with $\alpha_{initial}=\pm0.5$ along the $\hat{X}$ and $\hat{P}$ axes. The corresponding simulated and experimental results are presented in the Supplementary Material. 

\section{THE EXPERIMENT}

The CPG protocol was experimentally demonstrated on a single $^{171}\text{Yb}^+$ ion confined in a linear Paul trap. The $^2S_{1/2}$ hyperfine `clock' states, $\ket{g} \equiv \ket{F=0,m_{F}=0}$ and $\ket{e} \equiv \ket{F=1,m_{F}=0}$, form an effective two-level qubit. The trap has one axial ($Ax$) and two radial ($R1$ and $R2$) motional modes, with frequencies $\{{\omega_{Ax},\omega_{R1},\omega_{R2}}\}\approx 2\pi\times\{{0.62,1.01,1.35}\}$ MHz. Here, we demonstrate the protocol on the $R1$ mode. The calibration procedure of our SDF pulses follows \cite{tingrei2023calibrate} and is detailed in our Supplementary Material. 

The protocol starts with the qubit state initialized to $\ket{g}$ and the motional mode is cooled to $\bar{n} < 0.05$, where $\bar{n}$ is the average occupation number of the oscillator. From this point on, all operations carried out before qubit state readout were stitched together and implemented in a single, uninterrupted Arbitrary Waveform Generator (AWG) pulse. This AWG pulse first drives our Raman lasers to implement a $\pi/2$ carrier rotation to bring our qubit state to $\ket{-}$. When required, any non-vacuum initial bosonic states is then prepared. For each value of protocol length $N$, we program the set of gate angles and the pulse durations which we have numerically optimized in Section \ref{sec:Numerical Optimization}. Finally, we drive another $\pi/2$ carrier rotation to bring our qubit spin back to $\ket{g}$ in preparation for state tomography. The total duration of the protocol (including two carrier pulses of $25$ \si{\micro\second} each) is typically $250$ \si{\micro\second}, $400$ \si{\micro\second}, $400$ \si{\micro\second} for $N=6,10,14$ respectively. 

We perform state tomography by measuring the characteristic function ($\chi(\beta)=\langle\hat{D}(\beta)\rangle$ where the expectation value is calculated with respect to the motional state) following the procedure detailed in \cite{home2020}. As shown in Fig.~\ref{fig:circuit}, an initial $\pi/2$ carrier rotation is turned on to measure $\operatorname{Im}[\chi]$ or turned off to measure $\operatorname{Re}[\chi]$, while a SDF pulse selects the $\chi$ coordinate ($\beta$). The subsequent qubit bright state probability measurement yields a corresponding value of $\operatorname{Re}[\chi]$ or $\operatorname{Im}[\chi]$.    

More experimental details can be found in Appendix \ref{sec:experiment} and the Supplementary Material. 

\begin{figure*}[t]
\includegraphics[width=0.99\textwidth]{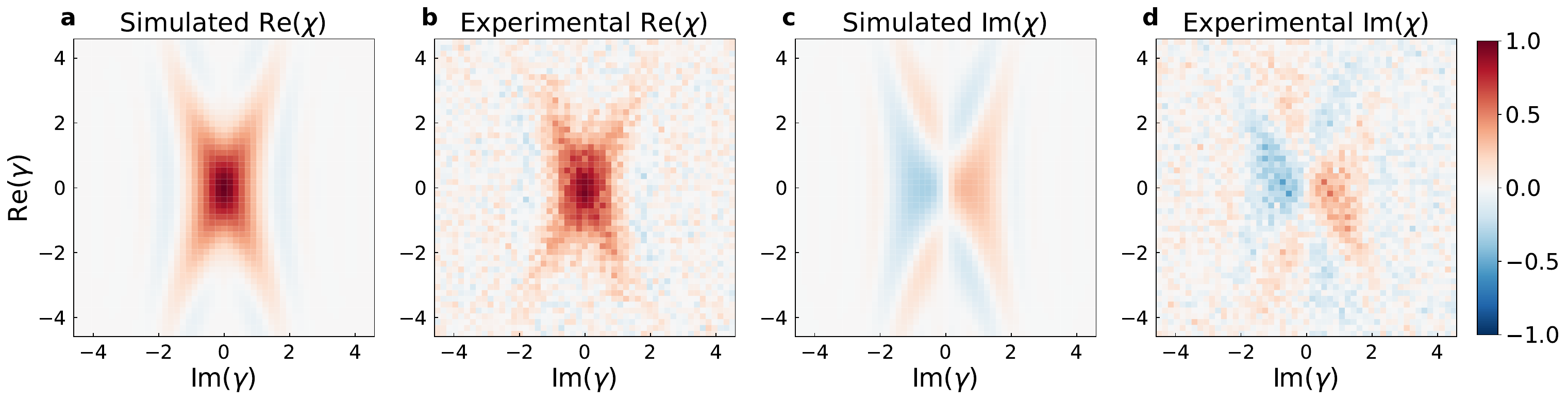}
\caption{(a) Simulated and (b) measured real component of the characteristic function after our $N=14$ CPG protocol. (c) Simulated and (d) measured imaginary component of the characteristic function after our $N=14$ CPG protocol. } 
\label{fig:14_pulse}
\end{figure*}

\begin{figure}[t] 
\includegraphics[width=0.49\textwidth]{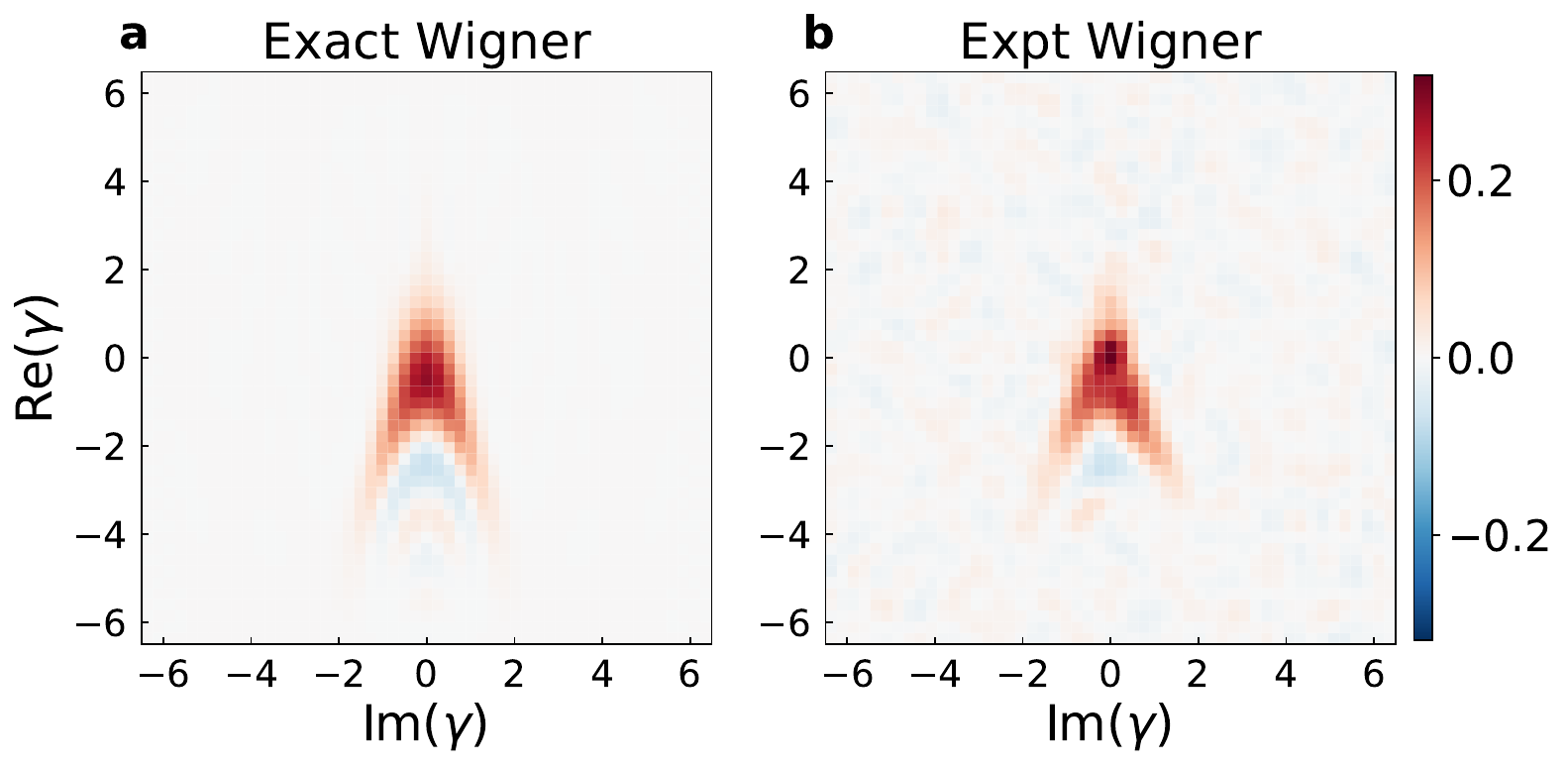}
\caption{(a) Simulated and (b) measured Wigner functions of an initial vacuum state after our $N=14$ CPG protocol. } 
\label{fig:expt_wigner}
\end{figure}

\begin{figure}[t] 
\includegraphics[width=0.45\textwidth]{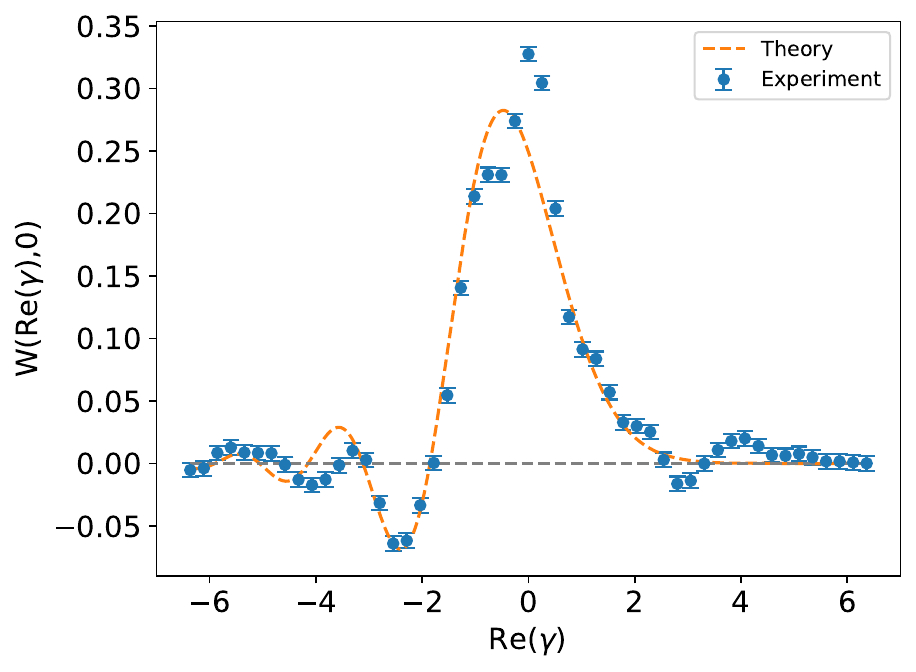}
\caption{Single slice of the 14-pulse CPG Wigner function at $\operatorname{Im}(\gamma)=0$. The blue circular markers represent experimental data with error bars calculated from projection noise over 200 experimental repetitions. The orange dotted line is generated from the same line cut of the Wigner function from the ideal cubic phase state.} 
\label{fig:wigner_slice}
\end{figure}

\section{RESULTS AND DISCUSSION} \label{sec:results}

To evaluate the performance of the protocol, we compare the state overlap between the Wigner function obtained via the cubic unitary evolution, the simulated protocol and the experiment. The Wigner function can be obtained from the characteristic function via the following expression \cite{home2020}: 

\begin{eqnarray}
 & & W(\gamma)=\frac{1}{\pi^2}\sum^\infty_{i,j=0}\chi(a_i,b_j)e^{\gamma\beta^*-\gamma^*\beta} \Delta a_i \Delta b_j.
\end{eqnarray}

Here, $a=\operatorname{Re}[\beta]$ and $b=\operatorname{Im}[\beta]$ are coordinates in characteristic function phase space, and $\{\Delta a_i, \Delta b_j\}$ refer to the interval between discrete steps in phase space. We choose $\Delta a_i = \Delta b_j = 0.18$ for our measurements and approximate the values of $\chi(\beta)$ outside our experimental range to be zero.

As shown in Figure \ref{fig:expt_wigner}, the experimental Wigner function shows good agreement with the simulated and ideal Wigner functions of the cubic phase state. As expected, the vertical slice of the Wigner function also matches the Airy function as shown in Figure \ref{fig:wigner_slice} \cite{ghose_non-gaussian_2007}. 

We calculate the state overlap via the discrete Wigner functions as follows:

\begin{eqnarray}
 & & F_{12} = \pi \sum^\infty_{i,j=0}  W_1(x_i,p_j)W_2(x_i,p_j) \Delta x_i \Delta p_j.
\end{eqnarray}

Here, $\{W_1,W_2\}$ refer to any two Wigner functions under comparison, $x\propto \operatorname{Re}[\gamma]$ and $p \propto \operatorname{Im}[\gamma]$ are coordinates on the Wigner function phase space, and $\{\Delta x_i, \Delta p_j\}$ refer to the interval in between discrete steps in phase space. We choose $\Delta x_i = \Delta p_j = 0.18 \times \sqrt{2}$ for our calculations and approximate the values of $W(\gamma)$ outside our experimental range to be zero.

We tabulate the various state overlaps measured between the ideal, simulated and experimental CPGs in Table \ref{tab:table1}. As expected, the 14-pulse protocol yielded the highest measured overlap. However, the overlap measured for the 10 pulse protocol was measured to be lower than the 6 pulse sequence. We attribute this to the fact that during the 10 pulse sequence, the bosonic mode reaches the highest maximum $\bar{n}$ compared to the 6 and 14 pulse protocol as shown in the Supplementary Material. At larger $\bar{n}$, undesirable effects from higher order terms in the Lamb-Dicke approximation and the dephasing of the bosonic mode become more significant, reducing CPG performance.  

For the same reasons, we observed lower overlaps (relative to the expected overlaps obtained from the simulation) when the motional state is initialized as a $|\alpha_{initial}|=0.5$ coherent state as opposed to the vacuum state. 

An alternative benchmark of the CPG performance is to measure the nonlinear squeezing (NLS) of our output state \cite{Moore_2019}. We first define a nonlinear quadrature $\hat{p}-3\lambda \hat{x}^{2}$ for cubic nonlinearity where $\lambda$ is some parameter, then we calculate the fluctuations of the nonlinear quadrature as a function of $\lambda$ as $V[\rho](\lambda) = \langle (\hat{p}-3\lambda \hat{x}^{2})^2\rangle-\langle \hat{p}-3\lambda \hat{x}^{2}\rangle^2$. These fluctuations of the nonlinear quadrature, also known as the NLS, for the cubic phase state is ideally suppressed below the nonlinear squeezing for the vacuum state.

\begin{eqnarray}
 & & V[\rho](\lambda)<V[\ket{0}\bra{0}](\lambda)
\end{eqnarray}

\begin{table}[tb]
\caption{\label{tab:table1}
Wigner overlaps measured for different values of initial coherent state amplitude $\alpha$ and protocol length $N$. Three different Wigner function overlaps are tabulated: overlaps between cubic unitary versus simulated $F_{US}$ cases, cubic unitary versus experimental $F_{UE}$ cases and simulated versus experimental $F_{SE}$ cases. Error bars are calculated from projection noise. 
}
\begin{ruledtabular}
\begin{tabular}{ccccc}
\textrm{$\alpha$}&
\textrm{$N$}&
\multicolumn{1}{c}{\textrm{$F_{US}$}}&
\textrm{$F_{UE}$}&
\textrm{$F_{SE}$}\\
\colrule
$0$ & 6 & 0.983 & 0.882(2)& 0.899(2)\\
$0$ & 10 & 0.995 & 0.864(2)& 0.868(2)\\
$0$ & 14 & 0.998 & 0.955(2)& 0.958(2)\\
$0.5$ & 14 & 0.980 & 0.908(2)& 0.928(2)\\
$-0.5$ & 14 & 0.980 & 0.844(2)& 0.871(2)\\
$0.5i$ & 14 & 0.998 & 0.930(2)& 0.931(2)\\
$-0.5i$ & 14 & 0.998 & 0.895(2)& 0.894(2)\\

\end{tabular}
\end{ruledtabular}
\end{table}

When this inequality is satisfied for some range of $\lambda$, nonlinear squeezing is observed and the cubic phase state possesses an advantage over the vacuum state as a quantum resource \cite{Moore_2019}. 

We first conduct two-dimensional fittings on the $N=14$ characteristic functions using our derived fitting functions (see Appendix \ref{appendix:derivations3} for details). NLS is then calculated by taking the higher-order derivatives of the fitted characteristic function about the origin \cite{gerry_introductory_2005} (please see Appendix \ref{appendix:derivations2} for details). Finally, we calculate the NLS uncertainty via Monte Carlo simulations where we randomly generate noise (modeled after projection noise) on the experimental data and do a re-fit on this ``noisy" dataset for every cycle. More details about the calculation of NLS can be found in the Supplementary Material. 

\begin{figure}[t] \label{fig:cubicity}
\includegraphics[width=0.45\textwidth]{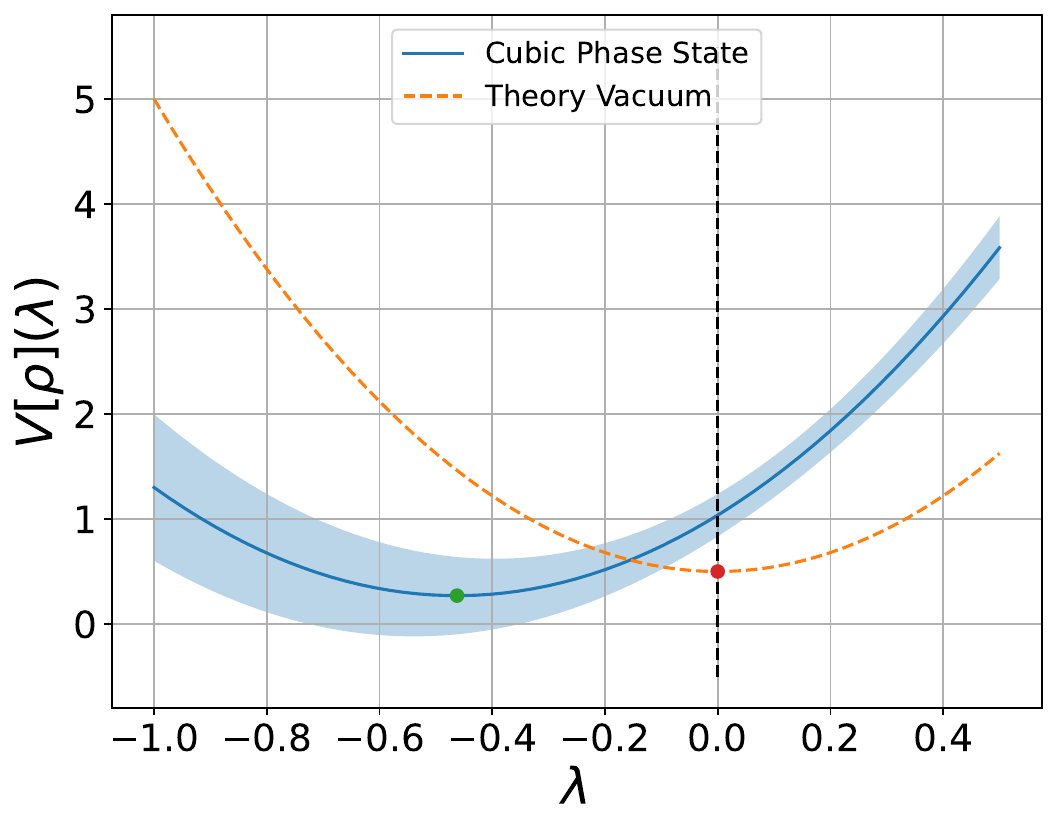}
\caption{Measured nonlinear squeezing (NLS) values as a function of the parameter ($\lambda$) calculated for a theoretical vacuum state (orange) and cubic phase state (blue). Here, the cubic phase state was prepared using our $N=14$ protocol. The point markers indicate the minimum point of the vacuum (red marker) and the cubic phase state (green marker) respectively. The blue shaded area represents one standard deviation of uncertainty that has been derived via the Monte Carlo method. Please refer to Appendix \ref{appendix:derivations2} and Supplementary Material for more details on the derivation of NLS values.} 
\label{fig:cubicity}
\end{figure}

As shown in Figure \ref{fig:cubicity}, taking into account the shaded area that represents one standard deviation of uncertainty, we observe suppression of the cubic phase state NLS below the vacuum for $\lambda \lesssim -0.23$.

\section{Concluding Remarks}

We have demonstrated a simple and efficient protocol capable of approximating the CPG on a trapped ion mechanical oscillator. The protocol only requires carrier qubit rotations and SDF pulses for both the CPG protocol and state tomography, which are both readily available on many trapped-ion platforms. The numerical optimization for the gate parameters can be further improved in the future with more sophisticated minimization protocols and addition of weights to each non-unique solution set. This workflow can be further extended to nonlinear gates beyond the CPG such as the quartic phase gate \cite{filip2024} and the double well potential \cite{tingrei2026}.

\section*{ACKNOWLEDGMENTS}

We would like to thank T. R. Tan, C. McGarry, K. Park and R. Filip for stimulating discussions and helpful comments. We would also like to thank J. You and R. Zhang for their constructive feedback and assistance. Finally, we thank ChatGPT 5.6 for its assistance in proofreading this manuscript.

This project is supported by the National Research Foundation, Singapore through the National Quantum Office, hosted in A*STAR, under its Centre for Quantum Technologies Funding Initiative (S24Q2d0009) and Quantum Engineering Programme (W21Qpd0208). 

\section*{Data Availability Statement}

The data that supports the findings of this article are available upon reasonable request from the authors.

\section*{AUTHOR CONTRIBUTIONS STATEMENT}

KMY, DM conceived the idea. KMY, NBLJ, DM developed the experimental protocols. NBLJ performed the experiment. NBLJ, EKJW analyzed the data. KMY, NBLJ, DM contributed to the experimental apparatus, and EKJW, DM provided theoretical support. DM supervised the project. NBLJ, EKJW, DM wrote the manuscript. All authors provided suggestions for the experiment, discussed the results, and contributed to the manuscript.

\appendix
\numberwithin{equation}{section}
\renewcommand{\theequation}{\thesection.\arabic{equation}}

\section{State-dependent force (SDF)} \label{appendix:appendix1}

The interaction Hamiltonian \cite{leibfried2003} between a trapped ion and a light field is written as 

\begin{align} \label{hamiltonian_JC}
\begin{split}
    H_{JC}(t) = & \frac{\hbar}{2}\Omega_0\hat{\sigma}_+(\mathds{1}+i\eta(\hat{a}e^{-i\nu t} + \hat{a}^\dagger e^{i\nu t})) e^{i(\phi-\delta t)} \\& + h.c.,
\end{split}
\end{align}

\noindent where $\Omega_0$ is the Rabi frequency, $\hat{\sigma}_\pm=(\hat{\sigma}_x\pm i\hat{\sigma}_y)/2$ are the qubit raising(+)/lowering(-) operator, $\eta$ is the Lamb-Dicke parameter, $\phi$ is the phase and $\delta$ is the laser detuning away from the motional mode frequency $\nu$. Here, we have taken the Lamb-Dicke approximation $(\eta^2(2\bar{n}+1))\ll 1)$. 

We control the qubit by setting the laser field detuning to be resonant with the qubit transition $(\delta=0)$. We call this the carrier transition and is described by the following interaction Hamiltonian under rotating-wave approximation: 

\begin{align}
    H_{carr} = \frac{\hbar}{2}\Omega_0(\hat{\sigma}_+e^{i\phi_{C}}+\hat{\sigma}_-e^{-i\phi_{C}}),
\end{align}

\noindent where $\phi_C$ is the phase of the carrier pulse. 

We are also able to drive coherent qubit-oscillator interactions by setting the laser detuning to be resonant with the frequency of the motional mode of our choosing $(\delta=\pm\nu)$. Setting $(\delta=+\nu)$ or $(\delta=-\nu)$ gives the blue-sideband (BSB) and red-sideband (RSB) interactions respectively: 

\begin{align}
    H_{BSB} = \frac{i\hbar}{2}\Omega(\hat{\sigma}_+ \hat{a}^\dagger e^{-i(\phi_{B}+\pi/2)} - \hat{\sigma}_- \hat{a}e^{i(\phi_{B}-\pi/2)}) \\
    H_{RSB} = \frac{i\hbar}{2}\Omega(\hat{\sigma}_+ \hat{a} e^{-i(\phi_{R}+\pi/2)} - \hat{\sigma}_- \hat{a}^\dagger e^{i(\phi_{R}-\pi/2)}),
\end{align}

\noindent where $\Omega=\eta\Omega_0$ is the Rabi frequency for the sideband transitions, and $\phi_B$ and $\phi_R$ are the relative phases of the lasers for the BSB and RSB respectively. 

We entangle the ion's spin and motion for the duration of our nonlinear gates to simulate anharmonicity to the bosonic mode via SDF pulses. SDF pulses in the context of trapped ions consist of applying a bichromatic laser pulse on our ion that drives both BSB and RSB interactions simultaneously. By tuning the Rabi frequencies of the RSB and BSB interactions to be equal ($\Omega_{RSB}=\Omega_{BSB}=\Omega$ when $n=0$) and the asymmetric detuning from carrier to zero ($|\delta|=|\nu|$), we obtain an interaction Hamiltonian for the bichromatic laser field under the rotating wave and Lamb-Dicke approximations \cite{monroe2018,SDF1,SDF2,SDF3,SDF4}. 

\begin{multline} \label{eq:H_I}
    H_I(t)= H_{BSB} + H_{RSB} \\
    = -\frac{\hbar}{2}\Omega(e^{-i\phi_s}\hat{\sigma}_{+}+e^{i\phi_s}\hat{\sigma}_-) \\\times(\hat{a}e^{i\delta t+i\phi_m} + \hat{a}^\dagger e^{-i\delta t-i\phi_m})
\end{multline}

The phase terms in the Hamiltonian $\phi_s$ and $\phi_m$ are known as the spin phase and motion phase terms, respectively, and are related to the phases of the RSB and BSB laser fields, where $\phi_s = (\phi_{B}+\phi_{R})/2$ and $\phi_m = (\phi_{B}-\phi_{R})/2$. By setting $\phi_m=0$, we find that this interaction Hamiltonian is exactly the SDF unitary operator (\ref{eq:SDF}) with $\phi = \phi_s$ \cite{monroe2018}. More details are available in the Supplementary Material.

\section{Calculation of nonlinear squeezing} \label{appendix:derivations2}

As discussed in Section \ref{sec:results}, we calculate the nonlinear squeezing (NLS) of our experimental cubic phase state as an alternative benchmark of CPG performance. Here, we show how we calculate the NLS from the characteristic function:

We first define the position and momentum operator:

\begin{align}
\begin{array}{cc}
    \hat{a} &= \frac{1}{\sqrt{2}}\qty(\hat{x}+i\hat{p})\\
    \hat{a}^\dagger &= \frac{1}{\sqrt{2}}\qty(\hat{x}-i\hat{p})
\end{array}
\quad \implies \quad
\begin{array}{cc}
    \hat{x} &= \frac{1}{\sqrt{2}}\qty(\hat{a}+\hat{a}^\dagger)\\
    \hat{p} &= \frac{1}{i\sqrt{2}}\qty(\hat{a}-\hat{a}^\dagger)
\end{array}
\end{align}

Next, we define $\alpha$ in terms of its real and imaginary parts:
\begin{align*}
    \beta = b_1+ib_2, \quad \{a,b\}\in \mathbb{R}. 
\end{align*}
The characteristic function can be re-written in the position and momentum operator as such:
\begin{align}
    \chi(b_1,b_2) &= \expval{\exp((b_1+ib_2) \hat{a}^\dagger-(b_1-ib_2))\hat{a})}_\rho\nonumber\\
    &=\expval{\exp( b_1(\hat{a}^\dagger-\hat{a})+ib_2(\hat{a}^\dagger+\hat{a}))}_\rho\nonumber\\
    &=\expval{\exp( i\sqrt{2}(b_2\hat{x}-b_1\hat{p}))}_\rho. 
\end{align}
We note that $V^3[\rho](\lambda)$ can be split into the following sum of components
\begin{align}
    V^3[\rho](\lambda) &= Var(\hat{p}_{NLQ})\nonumber\\
    &= \expval{(\hat{p}-3\lambda \hat{x}^2)^2}_\rho-\expval{\hat{p}-3\lambda \hat{x}^2}^2_\rho\nonumber\\
    &=\expval{\hat{p}^2}_\rho-3\lambda \expval{\hat{p}\hat{x}^2+\hat{x}^2\hat{p}}_\rho\nonumber\\
    &\phantom{\quad\quad\quad\quad}+9\lambda^2\expval{\hat{x}^4}_\rho-\qty(\expval{\hat{p}}_\rho-3\lambda\expval{\hat{x}^2}_\rho)^2. 
\end{align}
Using the property of characteristic functions, we can obtain the components as follows:
\begin{align}
    \expval{\hat{p}}_\rho &= -\frac{1}{i\sqrt{2}}\frac{\partial\chi(b_1,b_2)}{\partial b_1}\bigg\vert_{b_1,b_2 = 0}\\
    \expval{\hat{p}^2}_\rho &= -\frac{1}{2}\frac{\partial^2\chi(b_1,b_2)}{\partial b_1^2}\bigg\vert_{b_1,b_2 = 0}\\
    \expval{\hat{x}^2}_\rho &= -\frac{1}{2}\frac{\partial^2\chi(b_1,b_2)}{\partial b_2^2}\bigg\vert_{b_1,b_2 = 0}\\
    \expval{\hat{x}^4}_\rho &= \frac{1}{4}\frac{\partial^4\chi(b_1,b_2)}{\partial b_2^4}\bigg\vert_{b_1,b_2 = 0}\\
    \expval{\hat{p}\hat{x}^2+\hat{x}^2\hat{p}}_\rho&=\frac{1}{i\sqrt{2}}\frac{\partial^3\chi(b_1,b_2)}{\partial b_1\partial b_2^2}\bigg\vert_{b_1,b_2 = 0} \label{eq:cross-term}. 
\end{align}
Due to the existence of cross terms between $\hat{x}$ and $\hat{p}$, additional proof is required for the last relation \ref{eq:cross-term} and can be found in the Supplementary Material. 

\section{Two-dimensional fitting functions} \label{appendix:derivations3}
From Appendix \ref{appendix:derivations2}, calculating the nonlinear squeezing requires taking partial derivatives of the characteristic functions. Hence, it is instructive to perform a curve fit to the experiment data to obtain a best fit characteristic function for this purpose.\\

The fitting function for the real and imaginary parts were found to be:  
\begin{align}
    \Re(\chi(a,b)) &=\frac{1}{(c_1+c_2b_1^2)^{1/4}}\exp(\Xi_R)\cos(\Xi_I)\\
        \Im(\chi(a,b))&= \frac{1}{(c_1+c_2b_1^2)^{1/4}}\exp(\Xi_R)\sin(\Xi_I), 
\end{align}
with
\begin{align}
    \Xi_R &= \frac{c_3b_1^2+c_4b_2^2+c_5b_1^2b_2+c_6b_1^4}{c_1+c_2b_1^2}\\
    \Xi_I &= \frac{c_{7}b_1b_2^2+c_{8}b_1^3+c_{9}b_1^3b_2+c_{10}b_1^5}{c_1+c_2b_1^2}+c_{11}\tan^{-1}\qty(c_{12}b_1), 
\end{align}

\noindent where $\{c_1,\dots, c_{12}\}$ are fitting parameters. Details of the derivation can be found in the Supplementary Material.

\section{Experimental details}\label{sec:experiment}
The experimental platform consists of a linear radio-frequency (RF) Paul trap with four rods and two needle electrodes \cite{PhysRevA.76.052314}. An RF signal at 17.7 MHz is applied to one diagonal pair of rods and DC voltages are applied to all four rods and both needles. Within the Paul trap, we confine a single $^{171}\text{Yb}^+$ ion and encode quantum information within the hyperfine `clock' states $|g\rangle \equiv |F=0,m_{F}=0\rangle$ and $|e\rangle \equiv |F=1,m_{F}=0\rangle$. These hyperfine qubit states lie within the $^{2}\text{S}_{1/2}$ electronic ground state and are separated by 12.6 GHz \cite{PhysRevA.76.052314}. The transition between $^2S_{1/2}$ and $^2P_{1/2}$ at 369.5 nm is used to perform Doppler \cite{leibfried2003} and electromagnetically-induced transparency (EIT) cooling \cite{PhysRevLett.125.053001}, optical pumping, and state detection. 

Coherent stimulated Raman transitions between the qubit states are implemented using a 355 nm mode-locked pulsed laser split into two beams. Each arm is passed through an acousto-optic modulator (AOM), and the first-order diffracted beams are sent to the Paul trap. The AOMs allow us to control the frequency, amplitude, phase, and duration of the beams. The two Raman beams propagate into the vacuum chamber with an angle of $135^{\circ}$ relative to each other. The beat note between the two Raman beams drives coherent rotations within our qubit Bloch sphere.

An experimental run begins with the ion cooled to near the motional ground state ($\bar{n} < 0.05$) using Doppler cooling \cite{leibfried2003}, EIT cooling \cite{PhysRevLett.125.053001}, and resolved sideband cooling \cite{leibfried2003}. Doppler cooling is conducted with a 369.5 \si{\nano\meter} laser that is red-detuned by $15$ \si{\mega\hertz} from the $^{171}\text{Yb}^+$ ion's $\prescript{2}{}{S}_{1/2} \rightarrow \prescript{2}{}{P}_{1/2}$ transition. EIT cooling is conducted via a pair of counter-propagating 369.5 \si{\nano\meter} beams, where one beam is detuned 56 \si{\mega\hertz} away from $\prescript{2}{}{S}_{1/2} \rightarrow \prescript{2}{}{P}_{1/2}$ resonance and the other beam is detuned 65.15 \si{\mega\hertz} away from the same resonance. Pulsed sideband cooling is conducted on all three modes of the ion using our 355 \si{\nano\meter} Raman lasers. 

The ion spin is optically pumped to the $\ket{g}$ state via the $\prescript{2}{}{S}_{1/2} \ket{F=1} \rightarrow \prescript{2}{}{P}_{1/2} \ket{F=1}$ transition. The state $\ket{-} = (\ket{g} - \ket{e})/\sqrt{2}$ is then prepared using a slow $\pi/2$ rotation pulse ($\tau_{\pi/2} \approx 25 \; $\si{\micro\second}) resonant to the carrier transition.

To apply state-dependent forces (SDF) we illuminate the ion with bichromatic laser beams that simultaneously drive the RSB and BSB. For this experiment, we implement all SDF pulses on a radial mode with a mode frequency of approximately 1.01 MHz. 

We measure the population of the qubit state, corresponding to projections to the $Z$ eigen-states. We follow the procedure laid out in \cite{home2020} to measure our characteristic function $\chi(\beta)$. Since $\chi(\beta)$ is a complex number, we measure the real and imaginary components separately. The protocol used to measure the imaginary component consists of applying a $\pi/2$ carrier pulse followed by an SDF pulse which has a specific amplitude and motional phase $(\phi_m)$ that determines the co-ordinate $\beta$. To measure the real component, we remove the $\pi/2$ carrier pulse. 

After these secondary rotations, a state detection beam illuminates the ion for $ 1$ ms. The scattered photons are collected and relayed onto a photomultiplier tube (PMT). This allows for detection of the ion's excited state population, where a collection of $<2$ photons indicates a qubit in $\ket{g}$, and $\ket{e}$ otherwise. 

The qubit bright-state measurement error in our setup is $2.0(1)\%$, limited mostly by the ion fluorescence collection efficiency, and the dark-state measurement error (to the $\ket{g}$ qubit state) is $1.3(1)\%$. The heating rate for our three motional modes are $\{7.0(4), 3.0(2), 5.8(3)\}$ phonons per second for the modes $\{Ax, R1, R2\}$ respectively. The estimated $1/e$ coherence times for our three motional modes are $\{49(7), 47(7), 60(20)\}$ \si{ms} for the modes $\{Ax, R1, R2\}$ respectively. The estimated coherence times were calculated via an exponential-fit extrapolation of a Ramsey-type measurement ranging from 0 \si{\milli\second} to 10 \si{\milli\second}. More details are available in the Supplementary Material.  

All numerical simulations and optimizations were performed using the QuTiP quantum computing framework with a Hilbert space dimension set to 100.

\clearpage 
\onecolumngrid 
\makeatletter
\counterwithout{equation}{section} 
\def\@hangfrom@section#1#2#3{#1#2. {\MakeUppercase{#3}}}
\widetext
\begin{center}
\textbf{\large Supplemental Material for: Cubic Phase Gate with a Trapped Ion Oscillator}
\end{center}

\setcounter{equation}{0}
\setcounter{figure}{0}
\setcounter{table}{0}
\setcounter{page}{1}
\setcounter{section}{0}

\makeatletter
\renewcommand{\theHsection}{S-\arabic{section}}
\renewcommand{\theHequation}{S-\arabic{equation}}
\renewcommand{\theHfigure}{S-\arabic{figure}}
\renewcommand{\theHtable}{S-\arabic{table}}

\renewcommand{\theequation}{S\arabic{equation}}
\renewcommand{\thefigure}{S\arabic{figure}}
\renewcommand{\thetable}{S\arabic{table}}
\renewcommand{\thesection}{\Roman{section}}
\makeatother
\renewcommand{\appendixname}{}

\setlength\parindent{0pt}

\section{State-dependent force}

Here, we derive the state-dependent force produced by simultaneous application of the first-order red- and blue-sidebands.

\begin{align*}
    S\hat{D}F(\phi, \theta) = \exp(i\sqrt{2}\xi\sigma_{\pi/2,\phi}\hat{Q}(\theta)),
\end{align*}
where $\xi$ is the displacement amplitude, $\sigma_{\phi_1,\phi_2} \equiv \cos{\phi_1}\hat{\sigma}_z +\sin{\phi_1}\cos{\phi_2} \hat{\sigma}_x +\sin{\phi_1}\sin{\phi}_2\hat{\sigma}_y$ and $\hat{Q}(\theta) = \frac{\hat{a}e^{-i\theta}+\hat{a}^\dagger e^{i\theta}}{\sqrt{2}}$.

To begin, we shall define the necessary expressions:
\begin{align*}
    H_{RSB} &= -\frac{\hbar\Omega}{2}\qty(\hat{\sigma}_+\hat{a}e^{-i\phi_R}+\hat{\sigma}_-\hat{a}^{\dagger}e^{i\phi_R})\\
    H_{BSB} &= -\frac{\hbar\Omega}{2}\qty(\hat{\sigma}_+\hat{a}^\dagger e^{-i\phi_B}+\hat{\sigma}_-\hat{a}e^{i\phi_B})\\
    \phi &= \frac{\phi_R+\phi_B}{2}\\
    \theta &= \frac{\phi_R-\phi_B}{2}\\
    \hat{\sigma}_\pm &= \frac{\hat{\sigma}_x\pm i\hat{\sigma}_y}{2}.
\end{align*}
Putting everything together,
\begin{align*}
    H_{I} &= H_{RSB}+H_{BSB}\\
    &=-\frac{\hbar\Omega}{2}\qty(\hat{\sigma}_+\hat{a}e^{-i\phi_R}+\hat{\sigma}_-\hat{a}^{\dagger}e^{i\phi_R}+\hat{\sigma}_+\hat{a}^\dagger e^{-i\phi_B}+\hat{\sigma}_-\hat{a}e^{i\phi_B})\\
    &=-\frac{\hbar\Omega}{2}\qty(\hat{\sigma}_+\qty[\hat{a}e^{-i\phi_R}+\hat{a}^\dagger e^{-i\phi_B}]+\hat{\sigma}_-\qty[\hat{a}^\dagger e^{i\phi_R}+\hat{a}e^{i\phi_B}])\\
    &=-\frac{\hbar\Omega}{2}\qty(\hat{\sigma}_+e^{-i\phi}\qty[\hat{a}e^{-i\theta}+\hat{a}^\dagger e^{i\theta}]+\hat{\sigma}_-e^{i\phi}\qty[\hat{a}^\dagger e^{i\theta}+\hat{a}e^{-i\theta}])\\
    &=-\frac{\hbar\Omega}{2}\qty(\hat{\sigma}_+e^{-i\phi}+\hat{\sigma}_-e^{i\phi})\qty[\hat{a}e^{-i\theta}+\hat{a}^\dagger e^{i\theta}]\\
    &=-\frac{\hbar\Omega}{4}\qty( (\hat{\sigma}_x+ i\hat{\sigma}_y)e^{-i\phi}+( \hat{\sigma}_x- i\hat{\sigma}_y)e^{i\phi})\qty[\hat{a}e^{-i\theta}+\hat{a}^\dagger e^{i\theta}]\\
    &= -\frac{\hbar\Omega}{2}\qty( \cos\phi\hat{\sigma}_x+\sin\phi\hat{\sigma}_y)\qty[\hat{a}e^{-i\theta}+\hat{a}^\dagger e^{i\theta}]\\
    &=-\frac{1}{\sqrt{2}}\hbar\Omega\sigma_{\pi/2,\phi}\hat{Q}(\theta).
\end{align*}
Therefore, 
\begin{align*}
    S\hat{D}F(\phi, \theta) &= \exp(-\frac{i}{\hbar}H_{I}t)\\
    &=\exp\left(\frac{i}{\sqrt{2}}\Omega t\sigma_{\pi/2,\phi}\hat{Q}(\theta)\right)\\
    &=\exp\left(i\sqrt{2}\xi\sigma_{\pi/2,\phi}\hat{Q}(\theta)\right),
\end{align*}
where we have re-labelled $\frac{1}{2}\Omega t\rightarrow \xi$ at the last equality.

\vspace{5mm}

The value of $\phi$ in the $\sigma_{\pi/2,\phi}$ operator characterizes the coupling between the ion spin and motion. $\phi$ can be tuned by adjusting the sum of $\phi_R$ and $\phi_B$ in $H_I$. A more general control of $\{\phi_1,\phi_2\}$ in $\sigma_{\phi_1,\phi_2}$ can be achieved by direct rotations of the qubit spin \cite{filip2024}.

\section{Numerically Optimised Cubic Phase Gate Angles} \label{angles}

Here we list the set of angles (in radians), $\phi_i$, and displacement amplitudes, $\xi_i$, we have obtained via numerical optimization for $N=6,10,14$ conditional displacement pulses. The gate angles and displacement amplitude pairs were applied in numerical sequence. Namely, the displacement pulse described with $\phi_0$ and $\xi_0$ was applied first, followed by $\phi_1$ and $\xi_1$ and so on. The gate angles $\phi_i$ refer to the spin phase angles $\phi_s$. \\

\begin{align} \label{eqn:6_pulse}
    \phi_0 = 1.197257 \quad \xi_0 = 0.577288, \quad \phi_1 = 3.422101 \quad \xi_1 = 0.548726 \nonumber \\
    \phi_2 = 4.684920 \quad \xi_2 = 0.541296, \quad \phi_3 = 3.422101 \quad \xi_3 = 0.541296 \nonumber \\
    \phi_4 = 1.197257 \quad \xi_4 = 0.548726, \quad \phi_5 = 0.000000 \quad \xi_5 = 0.577288 \nonumber \\
\end{align}

\begin{align} \label{eqn:10_pulse}
    \phi_0 = 1.330524 \quad \xi_0 = 0.577673, \quad \phi_1 = 3.375207 \quad \xi_1 = 0.588981 \nonumber \\
    \phi_2 = 4.116582 \quad \xi_2 = 0.702366, \quad \phi_3 = 5.962227 \quad \xi_3 = 0.535096 \nonumber\\
    \phi_4 = 1.598925 \quad \xi_4 = 0.504176, \quad \phi_5 = 5.962227 \quad \xi_5 = 0.504176 \nonumber \\
    \phi_6 = 4.116582 \quad \xi_6 = 0.535096, \quad \phi_7 = 3.375207 \quad \xi_7 = 0.702366 \nonumber \\
    \phi_8 = 1.330524 \quad \xi_8 = 0.588981, \quad \phi_9 = 0.000000 \quad \xi_9 = 0.577673 \nonumber \\
\end{align}

\begin{align} \label{eqn:14_pulse}
    \phi_0 &= 0.765980 \;\quad \xi_0 = 0.324970, \quad \phi_1 = 1.927790 \;\quad \xi_1 = 0.346851 \nonumber \\
    \phi_2 &= 3.484556 \;\quad \xi_2 = 0.372704, \quad \phi_3 = 4.040196 \;\quad \xi_3 = 0.510346 \nonumber\\
    \phi_4 &= 4.863219 \;\quad \xi_4 = 0.404079, \quad \phi_5 = 0.443267 \;\quad \xi_5 = 0.320600 \nonumber \\
    \phi_6 &= 1.389142 \;\quad \xi_6 = 0.314735, \quad \phi_7 = 0.443267 \;\quad \xi_7 = 0.314735 \nonumber \\
    \phi_8 &= 4.863219 \;\quad \xi_8 = 0.320600, \quad \phi_9 = 4.040196 \;\quad \xi_9 = 0.404079 \nonumber \\
    \phi_{10} &= 3.484556 \quad \xi_{10} = 0.510346, \quad \phi_{11} = 1.927790 \quad \xi_{11} = 0.372704 \nonumber \\
    \phi_{12} &= 0.765980 \quad \xi_{12} = 0.346851, \quad \phi_{13} = 0.000000 \quad \xi_{13} = 0.324970, \nonumber \\
\end{align}

where (\ref{eqn:6_pulse}), (\ref{eqn:10_pulse}) and (\ref{eqn:14_pulse}) give the optimized sequences for N = 6, 10, and 14, respectively.

\vspace{5mm}

These sequences can alternatively be implemented by directly rotating the qubit spin state before each SDF pulse (where $\phi_s=\phi_m=0$). One can obtain these qubit rotation angles (in radians) by taking the difference between the current $\phi_i$ and the previous $\phi_{i-1}$. For example,\\

\begin{align*}
    \phi^{(spin)}_{0}&= \phi_0 \nonumber \\
    \phi^{(spin)}_{i}&= \phi_{i} - \phi_{i-1}, \nonumber \\
\end{align*}

where the first rotation is set to be the same angle as $\phi_0$ for all sequences. 

\newpage

\begin{figure}[h] 
\includegraphics[width=0.50\textwidth]{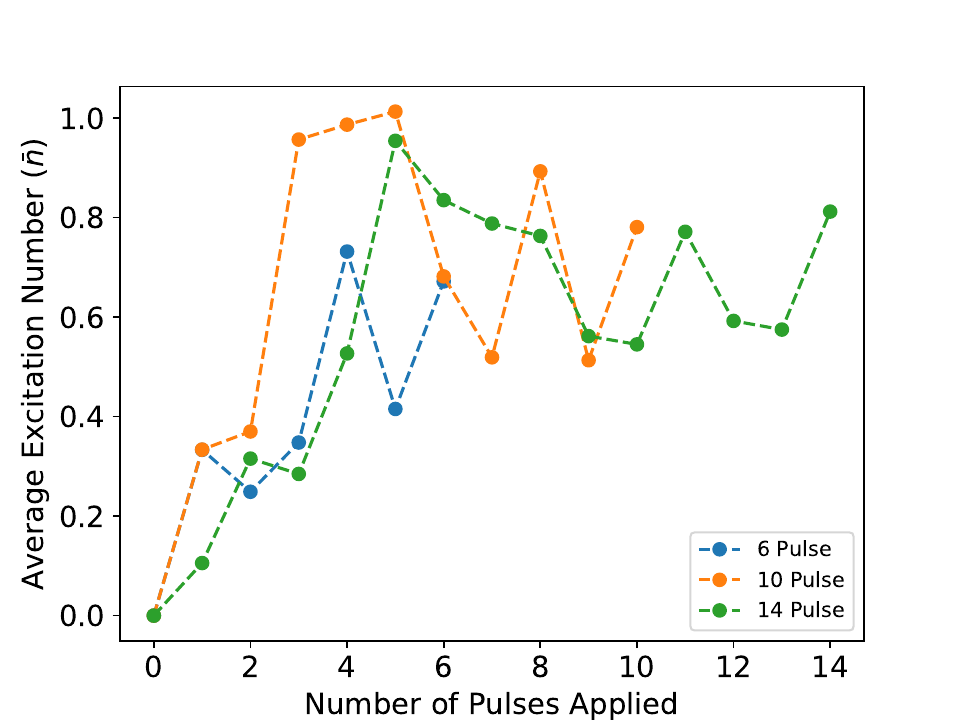}
\caption{Plot of the average excitation number of the intermediate bosonic states after a certain number of applied pulses. } 
\label{fig:nbar diagram}
\end{figure}

Using the optimized gate parameters stated above, we simulate the average excitation numbers of the intermediate bosonic states after every applied pulse as shown in Figure \ref{fig:nbar diagram}.

We also report the purities and the $\sigma_{x,y,z}$ components of the final qubit state after the various SDF pulse sequences and tracing out the oscillator state in Table \ref{tab:qubit}. Since we initialize the qubit in $\ket{-}$, the final qubit purities are close to $1$, the $\sigma_x$ components are close to $-1$, and the $\sigma_y$ and $\sigma_z$ components are close to $0$ as expected. These values also appear to approach their ideal values with increasing number of pulses, $N$. 

\begin{table}[h]
\caption{\label{tab:qubit}
Purities and $\sigma_{x,y,z}$ components of the qubit states after the $N=6,10,14$ CPG protocol are tabulated. With the assistance of QuTiP, we start in the vacuum state of harmonic oscillator and trace out the bosonic mode at the end of the various CPG sequences, leaving only the qubit state. Following which, we calculate purity via the trace, $\operatorname{tr}(\rho^2)$, and the various state components via taking expectations, $\langle\rho|\sigma_{x,y,z}|\rho\rangle$. 
}
\begin{ruledtabular}
\begin{tabular}{ccccc}
\textrm{$N$}&
\textrm{Purity}&

\multicolumn{1}{c}{\textrm{$\sigma_z$}}&
\textrm{$\sigma_y$}&
\textrm{$\sigma_x$}\\ [1ex]
\colrule \\ [0ex]
$6$ & 0.9930 & $0.0000$ & 0.0318& -0.9925\\ [1ex]
$10$ & 0.9976 & $0.0000$ & 0.0256& -0.9973\\ [1ex]
$14$ & 0.9992 & $0.0000$ & -0.0014& -0.9992\\ [1ex]

\end{tabular}
\end{ruledtabular}
\end{table}

\newpage
\section{Properties of state-dependent force sequences}
Here, we prove several useful properties of the state-dependent force decompositions.
\subsection{Symmetry of potential}
We first show that the sequences of state-dependent forces considered in this work can simulate only potentials containing odd powers of $\hat{X}$. Using the identity
$\hat{\sigma}_z \hat{\sigma}_\phi \hat{\sigma}_z = -\hat{\sigma}_\phi$,
we find that every state-dependent displacement pulse satisfies
\begin{equation}
\label{eq:sdf_conjugation}
\hat{\sigma}_z \mathrm{S\hat{D}F}(\phi,\xi)\hat{\sigma}_z
=
\hat{\sigma}_z
\exp\left(i\sqrt{2}\xi\hat{\sigma}_{\phi}\hat{X}\right)
\hat{\sigma}_z
=
\exp\left(-i\sqrt{2}\xi\hat{\sigma}_{\phi}\hat{X}\right).
\end{equation}
Therefore,
\begin{equation}
U(-\hat{X})=\hat{\sigma}_z U(\hat{X})\hat{\sigma}_z.
\end{equation}
Using the decomposition of $U(\hat{X})$ in terms of Pauli operators,
\begin{align}
\label{si:pauli}
U(-\hat{X})
&=
A(-\hat{X})\hat{I}
-i\left[
B_x(-\hat{X})\hat{\sigma}_x
+B_y(-\hat{X})\hat{\sigma}_y
+B_z(-\hat{X})\hat{\sigma}_z
\right]\nonumber
\\
&=
A(\hat{X})\hat{I}
-i\left[
-B_x(\hat{X})\hat{\sigma}_x
-B_y(\hat{X})\hat{\sigma}_y
+B_z(\hat{X})\hat{\sigma}_z
\right],
\end{align}
and comparing the two expressions, we obtain
\begin{align}
A(-\hat{X})   &= A(\hat{X}), \nonumber\\
B_x(-\hat{X}) &= -B_x(\hat{X}), \nonumber\\
B_y(-\hat{X}) &= -B_y(\hat{X}), \nonumber\\
B_z(-\hat{X}) &= B_z(\hat{X}).
\label{si:ab_symmetry}
\end{align}

Imposing the additional requirement that a qubit initially prepared in an eigenstate of the $\hat{\sigma}_x$ operator returns to the same state at the end of the sequence would ensure that $B_y(\hat X)=B_z(\hat X)=0$ and that the total unitary takes on the form:
\begin{equation}
U(\hat{X})
=
A(\hat{X})\hat{I}
-iB_x(\hat{X})\hat{\sigma}_x
=
\exp\left[-i\Phi(\hat{X})\hat{\sigma}_x\right],
\end{equation}
where $A(\hat{X})=\cos\left[\Phi(\hat{X})\right]$, $B_x(\hat{X})=\sin\left[\Phi(\hat{X})\right]$.
Given the symmetries of $A$ and $B_x$ established above, $\Phi(\hat{X})$ must be an odd function of $\hat{X}$:
\begin{equation}
\Phi(-\hat{X})=-\Phi(\hat{X}).
\end{equation}
More general forms of state-dependent forces, conditioned on the eigenstates of arbitrary Pauli operators, are required to simulate evolution in an arbitrary potential \cite{filip2024}.

\subsection{Palindromic sequences}

Let us consider a sequence of state-dependent forces in which both the amplitudes and the phases are palindromic: $\xi_i=\xi_{N-i-1}$, $\phi_i=\phi_{N-i-1}$.
Such a sequence has the form
\begin{equation}
U(\hat{X})
={}
\mathrm{S\hat{D}F}(\phi_0,\xi_0)
\mathrm{S\hat{D}F}(\phi_1,\xi_1)
\cdots
\mathrm{S\hat{D}F}(\phi_1,\xi_1)
\mathrm{S\hat{D}F}(\phi_0,\xi_0).
\end{equation}
For this sequence, $U^\dagger(\hat{X})=U(-\hat{X})$. Comparing the Pauli decomposition of $U^\dagger(\hat{X})$ with Eq.~\eqref{si:pauli}, we find that such a sequence satisfies $B_z(\hat{X})=0$ for every value of $\hat{X}$.

\subsection{Antipalindromic sequences}

Now consider a sequence in which the amplitudes are palindromic, $\xi_i=\xi_{N-i-1}$, whereas the phases are antipalindromic: $\phi_i=-\phi_{N-i-1}$. Such a sequence has the form
\begin{equation}
U(\hat{X}) = \mathrm{S\hat{D}F}(-\phi_0,\xi_0) \mathrm{S\hat{D}F}(-\phi_1,\xi_1)
\cdots
\mathrm{S\hat{D}F}(\phi_1,\xi_1) \mathrm{S\hat{D}F}(\phi_0,\xi_0).
\end{equation}
Because
$\hat{\sigma}_\phi^{T}=\hat{\sigma}_{-\phi}$, each pulse satisfies $ \mathrm{S\hat{D}F}(\phi,\xi)^T = \mathrm{S\hat{D}F}(-\phi,\xi)$. Transposition also reverses the order of the pulses. Together with the antipalindromic phase relation, this gives $ U(\hat{X})^T=U(\hat{X})$. Using the Pauli decomposition on both sides, we obtain
\begin{equation}
A(\hat{X})\hat{I}
-i\left[
B_x(\hat{X})\hat{\sigma}_x
-B_y(\hat{X})\hat{\sigma}_y
+B_z(\hat{X})\hat{\sigma}_z
\right]
=A(\hat{X})\hat{I}
-i\left[
B_x(\hat{X})\hat{\sigma}_x
+B_y(\hat{X})\hat{\sigma}_y
+B_z(\hat{X})\hat{\sigma}_z
\right].
\end{equation}
This equality requires $B_y(\hat{X})=0$.

Although we did not fully exploit these symmetries in the experimental sequences presented above, they may provide useful insights to construct optimal sequences in future work.

\subsection{Relation to the Fourier transform}
\label{SI:gate_properties}

Equation~(1) of the main text can also be expressed as a sum over spin paths. Introducing the projectors
\begin{equation}
\hat{P}_s(\phi)
=
\frac{\hat{I}+s\hat{\sigma}_{\phi}}{2},
\qquad
s=\pm 1,
\end{equation}
we can write the state-dependent force operator as
\begin{equation}
\mathrm{S\hat{D}F}(\phi_j,\xi_j)
=
\sum_{s=\pm 1}
\hat{P}_s(\phi_j)
\exp\left(is\sqrt{2}\xi_j\hat{X}\right).
\end{equation}
Substituting this expression into the full pulse sequence gives
\begin{equation}
\exp\left[i\Phi(\hat{X})\right]
=
\sum_{s_0,\ldots,s_{N-1}=\pm 1}
C_{\boldsymbol{s}}
\exp\left(i\sqrt{2}\lambda_{\boldsymbol{s}}\hat{X}\right),
\end{equation}
where $\boldsymbol{s}=(s_0,\ldots,s_{N-1})$,
$\lambda_{\boldsymbol{s}} = \sum_{k=0}^{N-1}s_k\xi_k$, and
$ C_{\boldsymbol{s}} = \langle - | \hat{P}_{s_{N-1}}(\phi_{N-1}) \cdots \hat{P}_{s_0}(\phi_0) | - \rangle$.

In the special case in which all displacement amplitudes are equal, $\xi_k=\xi$, the signed sum $ m=\sum_{k=1}^{N}s_k$ can take only even integer values $m=-N,-N+2,\ldots,N-2,N$. The expression therefore reduces to the finite Fourier series
\begin{equation}
\exp\left[i\Phi(\hat X)\right] =
\sum_{l=-N/2}^{N/2}
C_m \exp\left( 2\sqrt{2} i l\xi\hat X \right)
\end{equation}
if the qubit returns back to $|-\rangle$ state. This connection to a finite Fourier series was previously pointed out in \cite{filip2024}.

\section{Calibration Protocol}

The experimental Cubic Phase Gate requires several calibration steps. \\

Firstly, the frequencies and $\pi$ times for each of the carrier, red-sideband (RSB), and blue-sideband (BSB) transitions were calibrated individually. The $\pi$ times for the RSB and BSB transitions were matched by adjusting the amplitude of each frequency component until a $\pi$ time of $50\mu s$ was achieved for all three transitions. We then calibrate the frequency detunings and laser phases for the state-dependent force (SDF) following the procedure described in \cite{tingrei2023calibrate}. \\

The frequencies of the BSB ($f_{{BSB}}$) and RSB ($f_{{RSB}}$) transitions are parameterized in terms of the carrier frequency ($f_{{carr}}$), symmetric detuning ($\Delta_{{sym}}$), and asymmetric detuning ($\Delta_{{asym}}$).

\begin{align*}
    f_{BSB} = f_{carr} - \Delta_{sym} + \Delta_{asym} \\
    f_{RSB} = f_{carr} + \Delta_{sym} + \Delta_{asym} \\
\end{align*}

\begin{figure}[h] 
\includegraphics[width=0.60\textwidth]{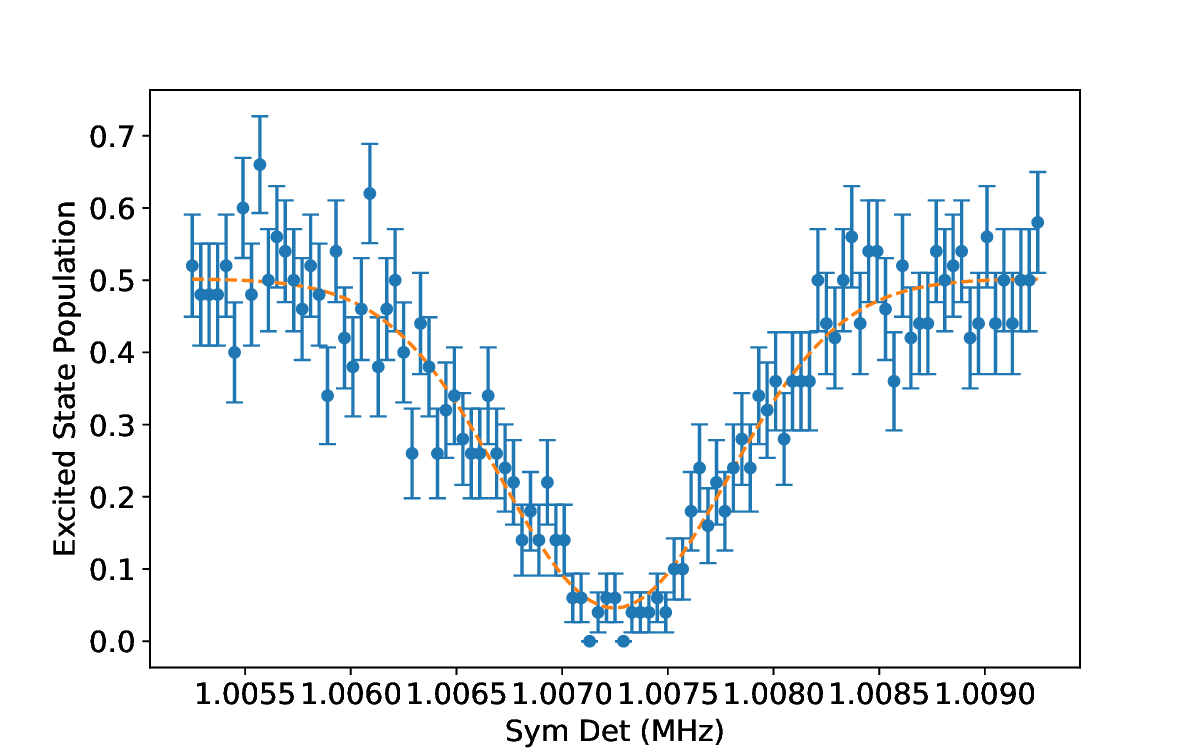}
\caption{Typical symmetric detuning calibration curve for the SDF pulse. A Gaussian function is used as the approximate fitting function (orange dashed line) for the data points (blue markers). The frequency where the fit reaches the minimum value is taken as the calibration result. In this example calibration curve, the optimal symmetric detuning is 1.00725(2)MHz.} 
\label{fig:sym det}
\end{figure}

\begin{figure}[h] 
\includegraphics[width=0.60\textwidth]{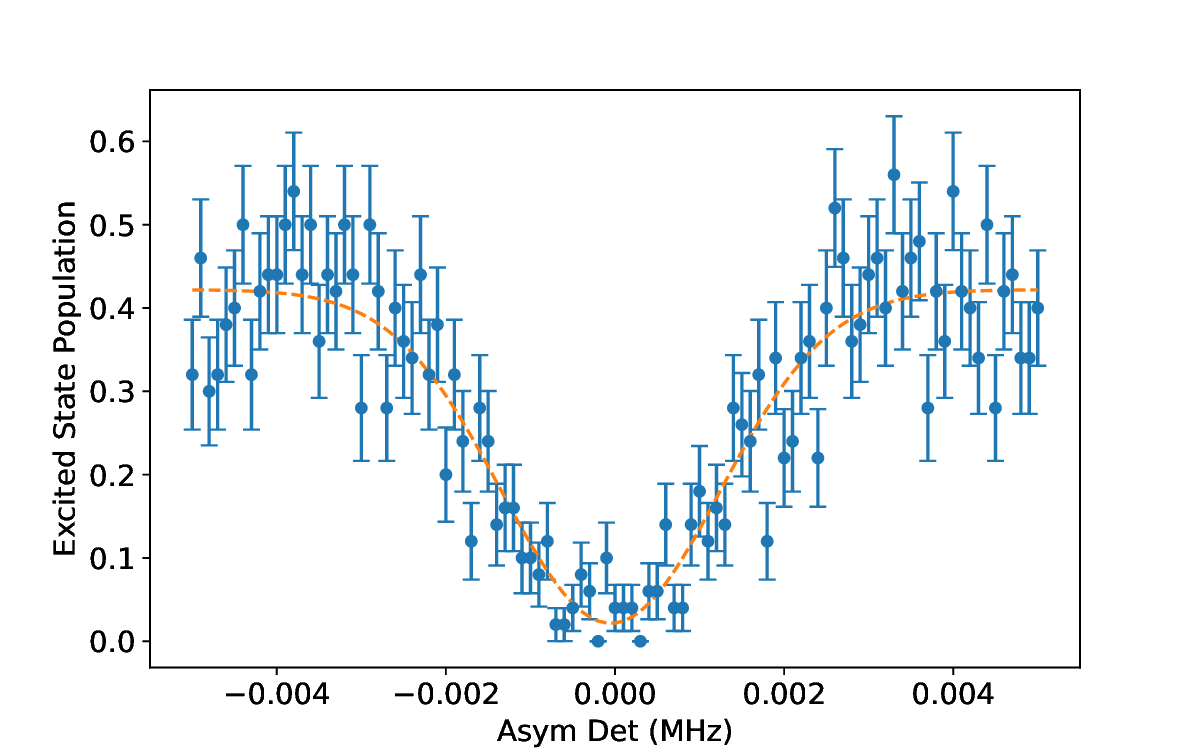}
\caption{Typical asymmetric detuning calibration curve for the SDF pulse. A Gaussian function is used as the approximate fitting function (orange dashed line) for the data points (blue markers). The frequency where the fit reaches the minimum value is taken as the calibration result. In this example calibration curve, the optimal asymmetric detuning is -50(60)Hz.} 
\label{fig:asym det}
\end{figure}

The frequency detunings can be calibrated using two consecutive SDF pulses where we set the phases for the first pulse as $\phi_s=\phi_m=0$ while the second pulse has the phases $\phi_s=\pi, \phi_m=0$. We arbitrarily set the duration of each SDF pulse to $100\mu s$. We then scan $\Delta_{{sym}}$ around its expected value, corresponding to the frequency of the addressed motional mode. The $\Delta_{sym}$ calibration curve is shown in Figure \ref{fig:sym det} where the calibrated result is the minimum of the dip. A similar procedure is used to calibrate $\Delta_{\mathrm{asym}}$, except that the detuning is scanned around zero \cite{tingrei2023calibrate} and a typical calibration curve is shown in Figure \ref{fig:asym det}. 

We calibrate the SDF pulse duration for any given SDF amplitude by measuring a one-dimensional slice of the real component of the characteristic function of a vacuum state along the line where $\operatorname{Re}[\beta]=0$ as a function of SDF pulse duration. The resulting characteristic-function slice is fitted to a Gaussian function (\ref{eq:gaussian}) where $A,\mu,\sigma,c$ denote the peak height, mean position, standard deviation, and offset values respectively. The width term $\sigma$ in the Gaussian fit function is associated with the extent of the vacuum-state wavepacket, whose characteristic length scale is $x_0\equiv\sqrt{\hbar/(2m\omega)}$, where $m$ is the mass of the ion and $\omega$ is the oscillation frequency of the corresponding motional mode. Following which, we take the fitted value of the standard deviation $\sigma$ in terms of SDF pulse duration and use this value as the definition of unity on the $\beta$ axis. 

\begin{align} \label{eq:gaussian}
    f(A,\mu,\sigma,c)=Ae^{-(\frac{x-\mu}{\sqrt{2}\sigma})^{2}}+c
\end{align}

The offset of the spin phase $\phi_s$ is calibrated by first preparing the qubit in the $\ket{+}$ state with a single-qubit rotation gate $R_y(\pi/2)$. An SDF pulse is then applied, after which the qubit is returned to the ground state by applying the inverse rotation $R_y^\dagger(\pi/2)$ \cite{tingrei2023calibrate}. The SDF pulse has a set duration that would enact an arbitrarily chosen displacement of $\xi = 1.50$. The spin phase $\phi_s$ is then scanned, and the resulting excited state population is fitted to a sinusoidal function. The calibrated value of $\phi_s=0$ can be chosen as any phase that results in a minimum of the fit function. For example, if a minimum occurs at $\phi_s=\pi/2$, this phase offset must be added to the optimized values of $\phi_s$ obtained in Section~\ref{angles}.

\begin{figure}[h] \label{fig:phase}
\includegraphics[width=0.49\textwidth]{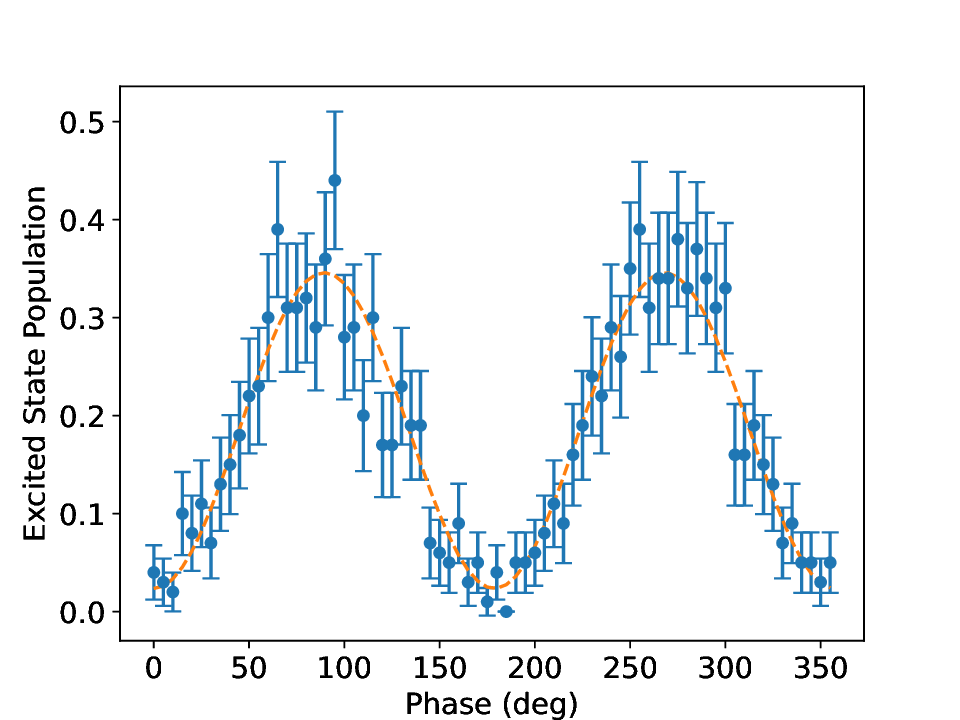}
\caption{Typical spin phase $(\phi_s)$ calibration curve for the SDF pulse. A $\sin^2$ function is used as the approximate fitting function (orange dashed line) for the data points (blue markers). The phase where the fit reaches the minimum value is taken as the calibration result. In this example calibration curve, the optimal phase is $-5(2)^\circ$.} 
\end{figure}

\newpage

\section{Characteristic Function Measurement}

We follow the methods laid out in \cite{home2020} to measure characteristic functions of motional states. \\

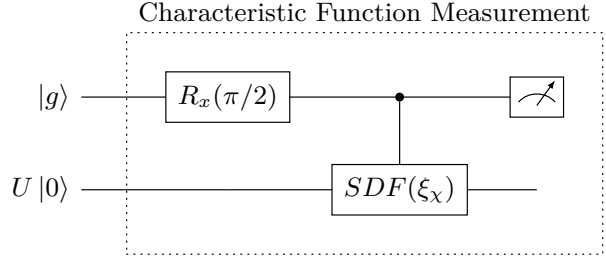
\begin{figure*}[h]
    \centering
    \resizebox{0.45\textwidth}{!}{
    \begin{quantikz}[thin lines,node distance=2pt] 
            \lstick{$\ket{g}$}& \qw& \gate{R_x(\pi/2)} \gategroup[wires=2,steps=3,style={dotted,cap=round,inner sep=10pt}]{Characteristic Function Measurement} & \ctrl{1} &\meter{} \\
            \lstick{$U\ket{0}$}& \qw & \qw &  \gate{SDF(\xi_\chi)} &\qw
    \end{quantikz}
    }
    \caption{Illustrated diagram of the characteristic function measurement circuit. It requires access to an ancillary qubit, which we assume starts off in ground state $\ket{g}$, and the chosen bosonic mode which has been subjected to an unknown unitary $U\ket{n=0}$ respectively. It consists of a single-qubit rotation gate ($R_x$) and another SDF pulse \cite{home2020}.}
    \label{fig:chi_circuit}
\end{figure*}

The $R_x(\pi/2)$ gate is turned off to measure $\operatorname{Re}[\chi]$ and turned on to measure $\operatorname{Im}[\chi]$. The ${R}$ gate is implemented with a Raman carrier pulse resonant to the qubit transition. The laser waveform phase for this rotation gate is set to have a $\pi/2$ phase difference relative to the CPG rotation $R_{y}$ gates used in the CPG protocol. \\

The SDF pulse in the measurement protocol is used to determine the coordinate on the $\chi(\beta)$ phase space being measured. We record our experimental measurements on a rectangular grid defined by the coordinates $\operatorname{Re}[\beta]$ and $\operatorname{Im}[\beta]$. However, the experimentally controlled parameters of the SDF pulse are the displacement amplitude $\xi$ and motional phase $\phi_m$, which provide a polar representation of the phase-space coordinates. We determine the SDF parameters using the following equations: \\

\begin{align} \label{eqn:coordinates}
    \xi_\chi &= \sqrt{\operatorname{Re}[\beta]^2 + \operatorname{Im}[\beta]^2} \\
    \phi_m&=\begin{cases}
			\pi/2, & \text{if $\operatorname{Im}[\beta] = 0$}\\
            \arctan{(\operatorname{Re}[\beta]/\operatorname{Im}[\beta])}, & \text{if $\operatorname{Im}[\beta] > 0$}\\
            \pi-\arctan{(\operatorname{Re}[\beta]/\operatorname{Im}[\beta])}, & \text{if $\operatorname{Im}[\beta] < 0$}.
		 \end{cases}\\ \nonumber
\end{align}

We measure $\chi(\beta)$ over the coordinate ranges $\operatorname{Re}[\beta]\in[0,4.5]$ and $\operatorname{Im}[\beta]\in[-4.5,4.5]$. We measure 51 points along the $\operatorname{Im}[\beta]$ axis and 26 points along the $\operatorname{Re}[\beta]$ axis. We then exploit the symmetry of $\chi(\beta)$ to extend the measured data to $\operatorname{Re}[\beta] \in [-4.5,4.5]$. Each pixel has an area of $0.18\times0.18$ and represents an average reading over 200 experimental repetitions. This average reading of the qubit excitation probability $(P_e)$ is used to calculate $\chi(\beta)$ via the scaling relation: \\

\begin{equation}
    \chi(\beta) = 2\times(P_e-1/2).
\end{equation}

We experimentally determine the background offset of $\chi(\beta)$ by measuring $\operatorname{Im}[\chi(\beta)]$ for the vacuum state, for which the ideal value is zero at all phase-space coordinates. We measure an average value of $-0.022$ and take this value to be subtracted from all future $\chi(\beta)$ measurements. This offset arises from the imbalance between the dark- and bright-state measurement errors characterized in Section~\ref{sec:errors}. 

\newpage

\section{Vacuum state and Coherent state measurement results}

\begin{figure}[h] 
\includegraphics[width=0.6\textwidth]{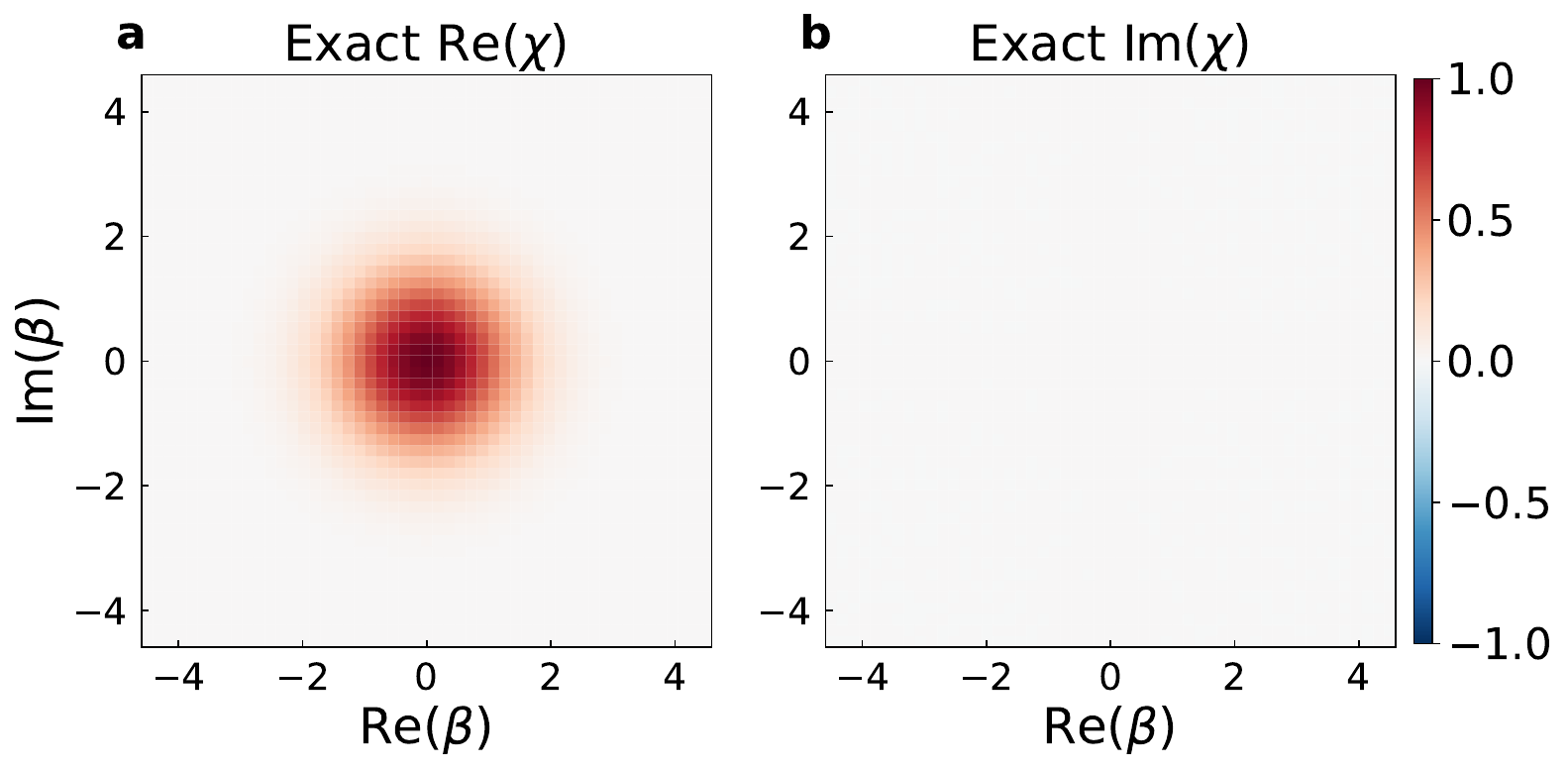}
\caption{Simulated real (a) and imaginary (b) components of the characteristic function describing a vacuum state.} 
\label{fig:vacuum thr chr}
\end{figure}

\begin{figure}[h]
\includegraphics[width=0.6\textwidth]{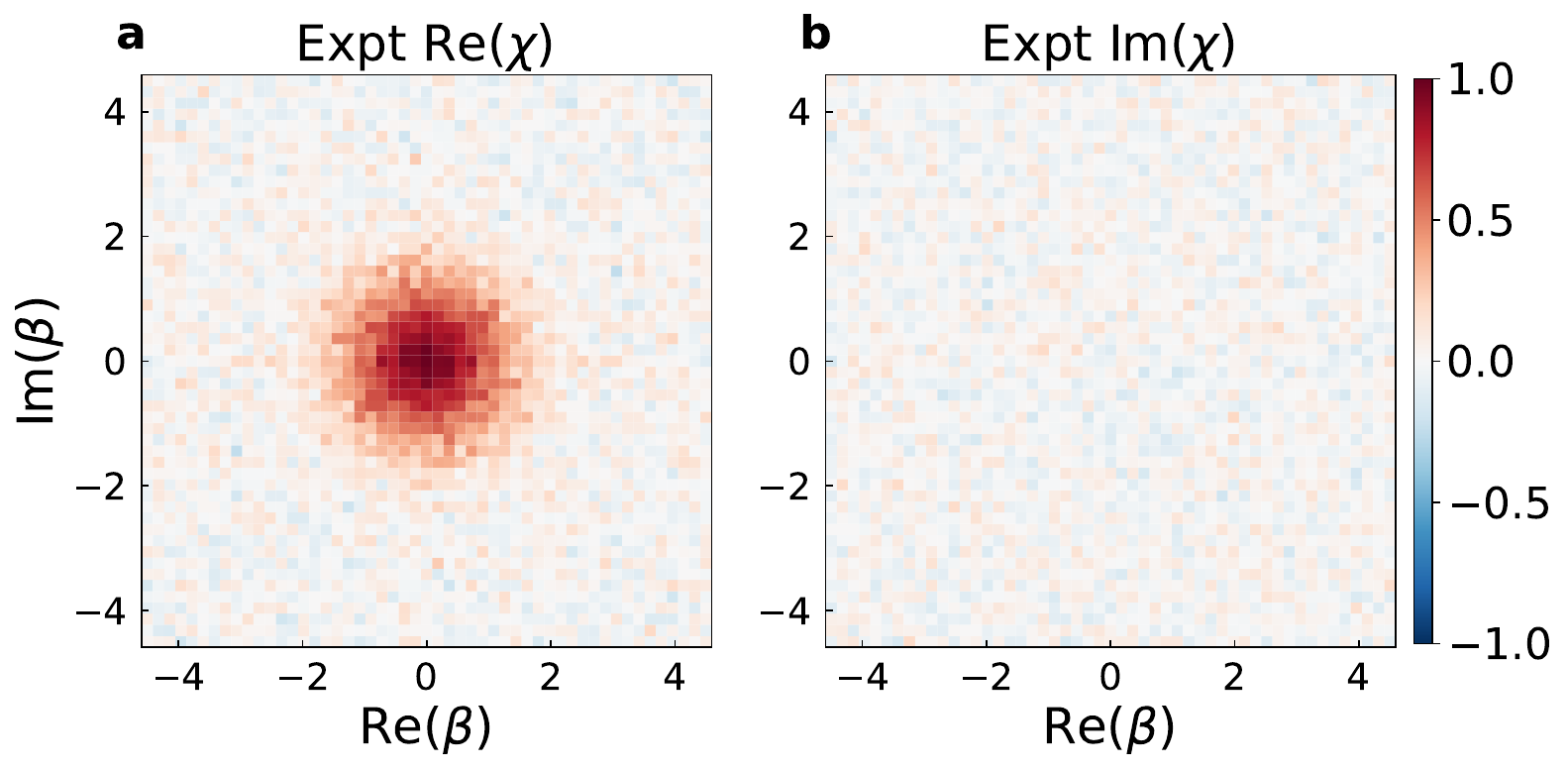}
\caption{Measured real (a) and imaginary (b) components of the characteristic function describing our experimental vacuum state.} 
\label{fig:vacuum expt chr}
\end{figure}

\begin{figure}[h]
\includegraphics[width=0.6\textwidth]{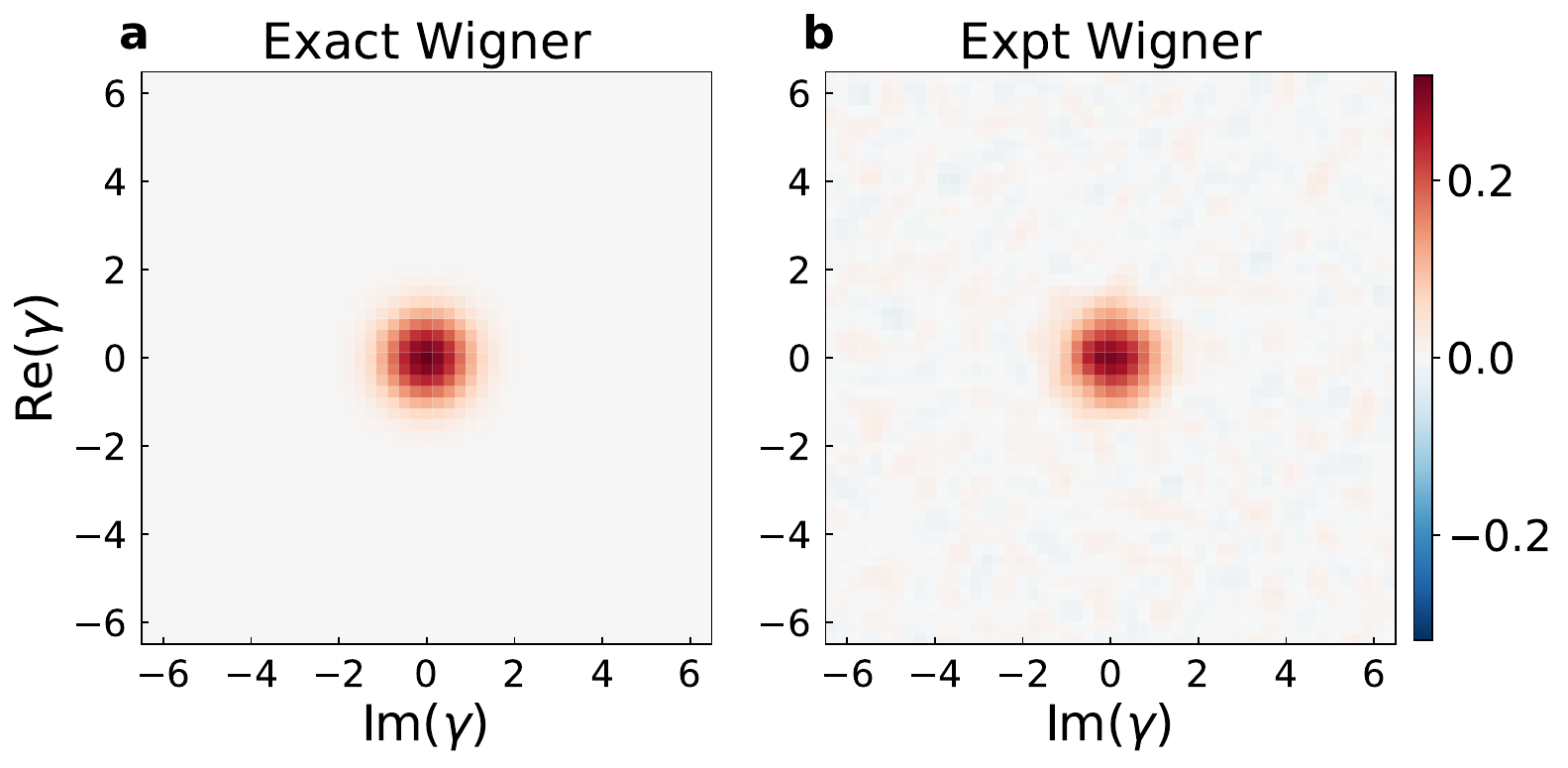}
\caption{Simulated (a) and experimental (b) Wigner functions describing our vacuum state.} 
\label{fig:vacuum wigner}
\end{figure}

\begin{figure}[h]
\includegraphics[width=0.6\textwidth]{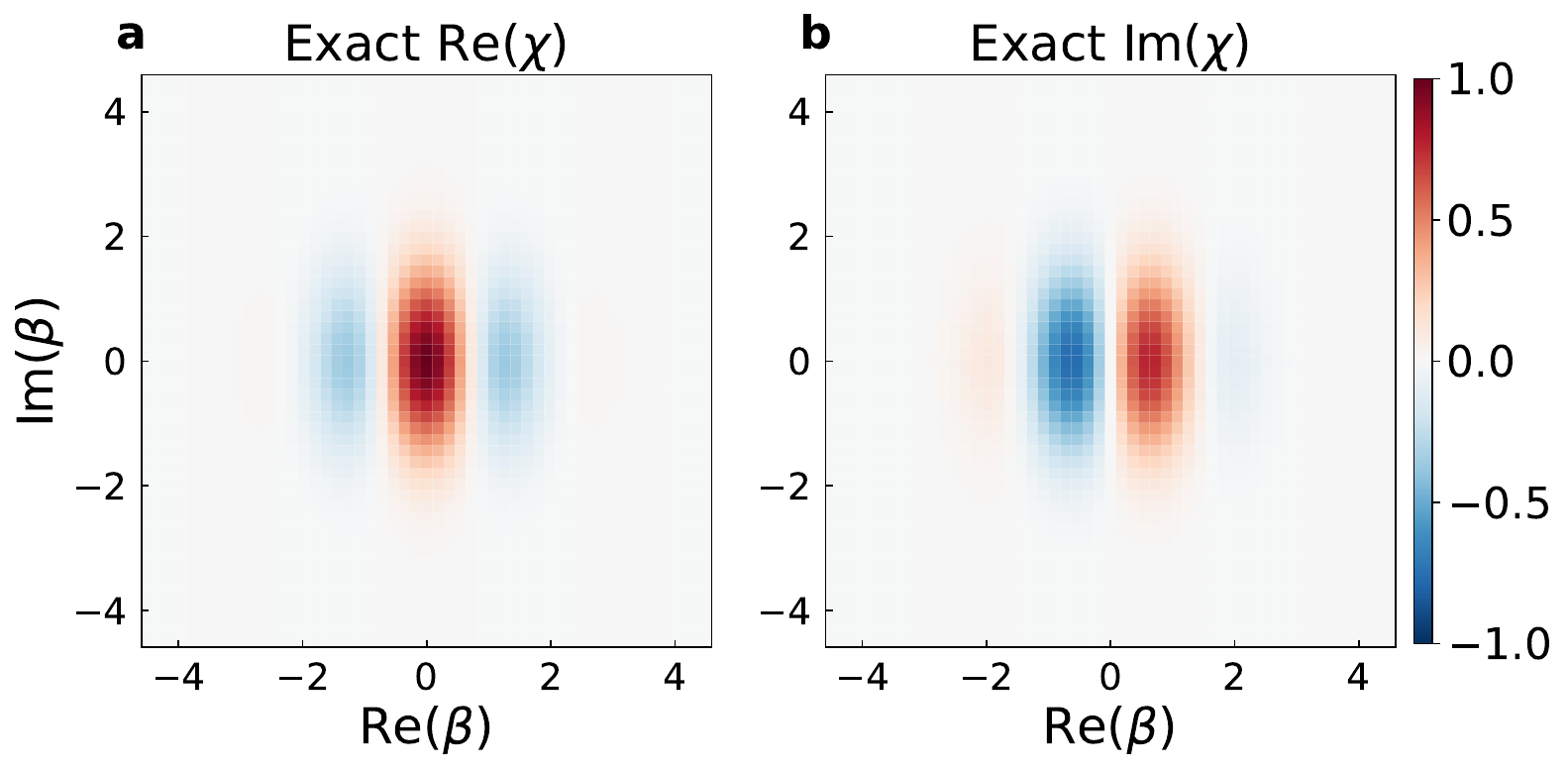}
\caption{Simulated real (a) and imaginary (b) components of the characteristic function describing a coherent $(\alpha=i)$ state.} 
\label{fig:coherent thr chr}
\end{figure}

\begin{figure}[h] 
\includegraphics[width=0.6\textwidth]{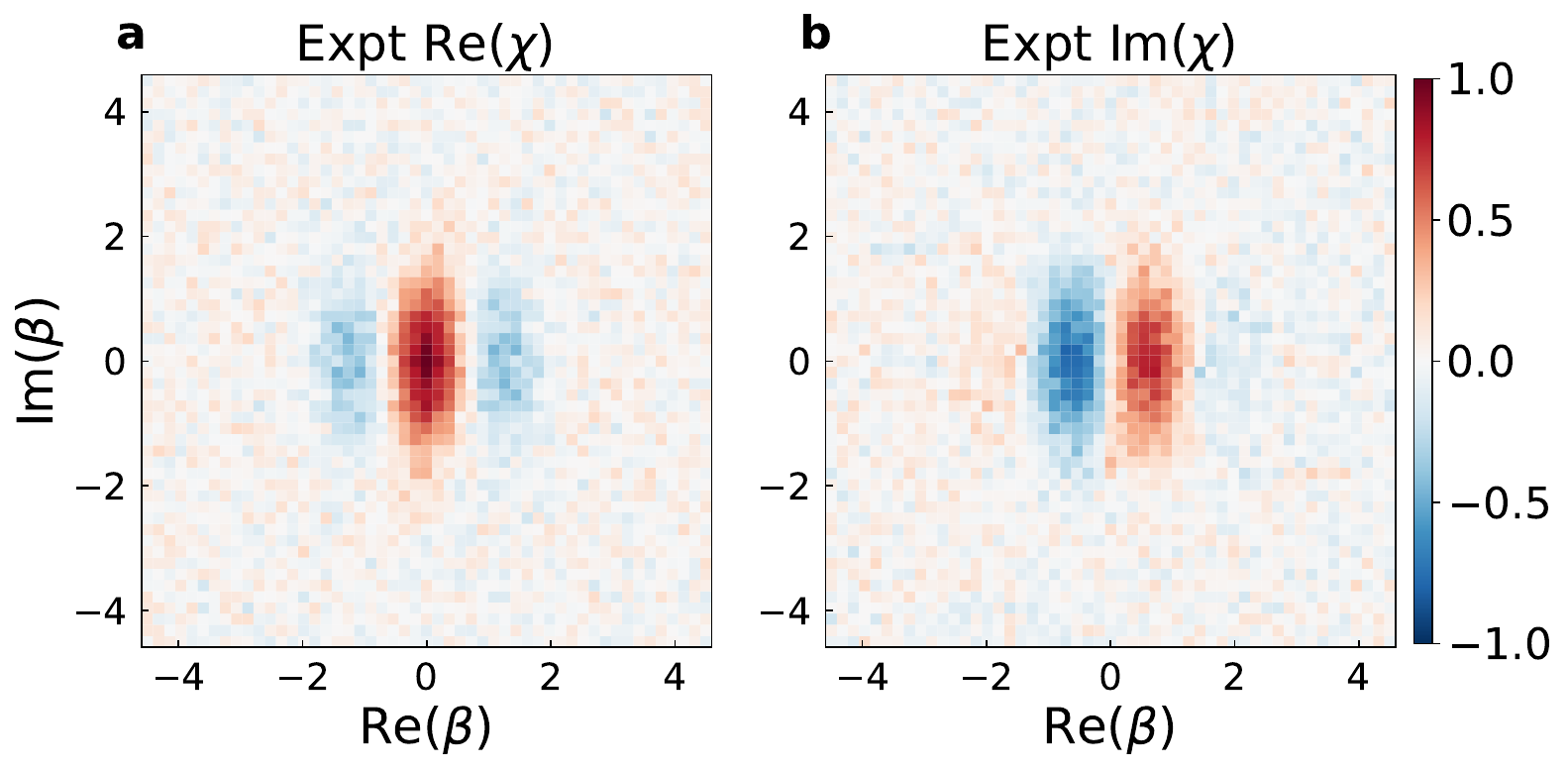}
\caption{Measured real (a) and imaginary (b) components of the characteristic function describing our experimental coherent $(\alpha=i)$ state.} 
\label{fig:coherent expt chr}
\end{figure}

\begin{figure}[h!] 
\includegraphics[width=0.6\textwidth]{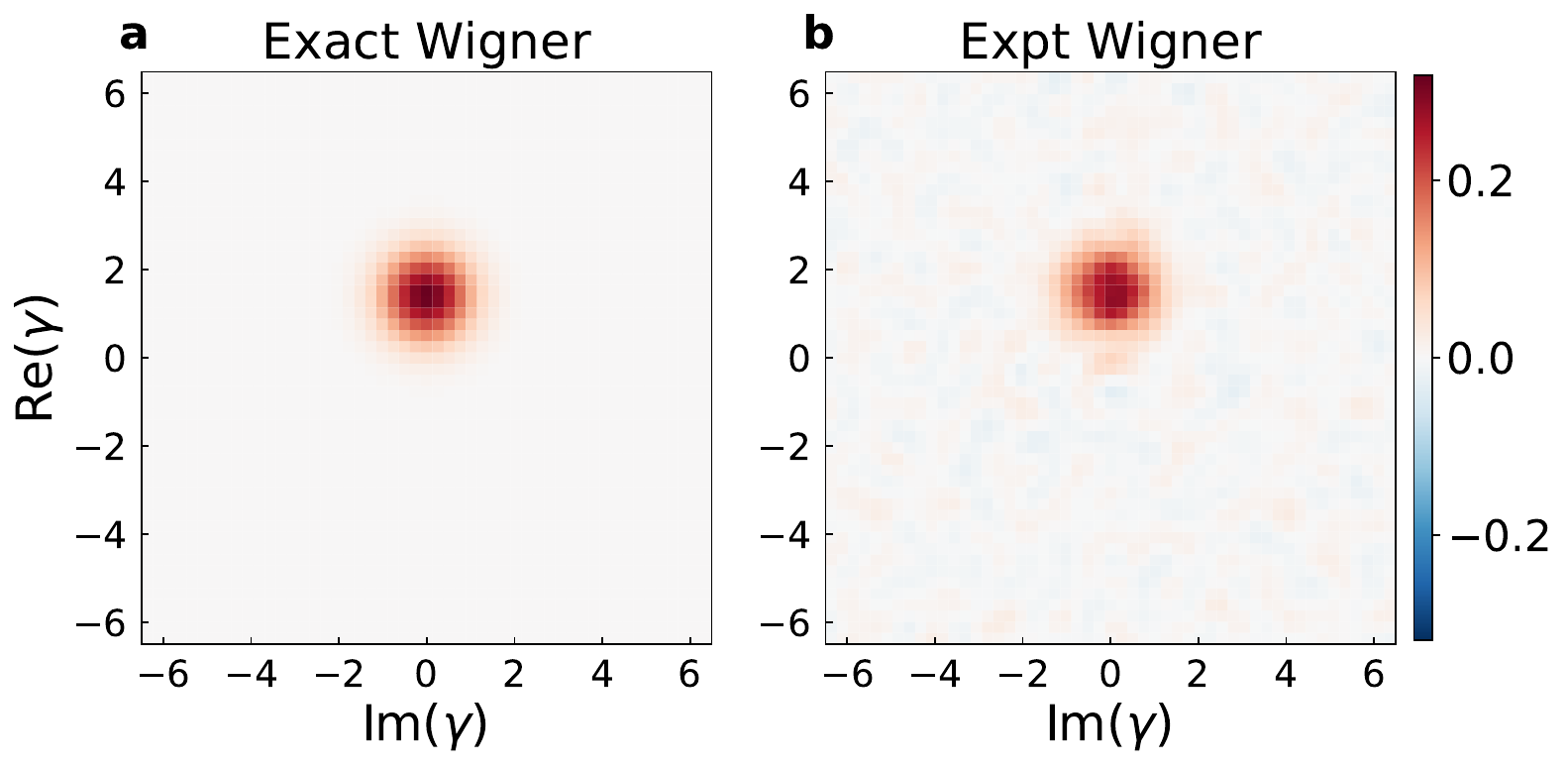}
\caption{Simulated (a) and experimental (b) Wigner functions describing our coherent $(\alpha=i )$ state.} 
\label{fig:coherent wigner}
\end{figure}

To calibrate and test our motional state preparation and measurement procedures, we measured the vacuum state and various coherent states. Figures \ref{fig:vacuum thr chr}, \ref{fig:vacuum expt chr}, \ref{fig:vacuum wigner} shows the simulated and measured results for both the characteristic functions and Wigner function for the vacuum state. Figures \ref{fig:coherent thr chr}, \ref{fig:coherent expt chr}, \ref{fig:coherent wigner} shows the same data taken for the $\alpha=i$ coherent state. 

\section{Motional Coherence, Motional Heating Rates and qubit SPAM errors} \label{sec:errors}

\begin{figure}[h] 
\colorbox{white}{
\includegraphics[width=0.99\textwidth]{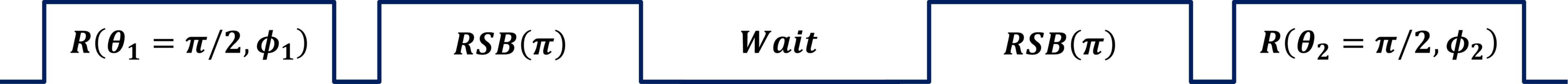}
}

\caption{Pulse sequence for measuring motional coherence time. We sandwich a variable delay with a pair of RSB $\pi$ pulses which in turn, is sandwiched between a pair of carrier rotation ($R$) $\pi/2$ pulses. After this Ramsey sequence, we conduct qubit state readout. We fix the phase of the first carrier pulse $\phi_1=0$ whilst scanning the phase of the second carrier pulse $\phi_2$ from $0$ to $2\pi$ to measure contrast. } 
\label{fig:coherence ramsey}
\end{figure}

We measure the motional coherence times of all three motional modes of a single trapped ion using a Ramsey-type measurement, as illustrated in Fig.~\ref{fig:coherence ramsey}. For each value of wait time from 0 to 10 ms, we measure its contrast by fitting the scan results to a sinusoidal function. As the motional mode decoheres over time, the measured contrast is expected to decay with increasing wait time. We plot a graph of contrast values against wait time to estimate the motional coherence. \\

\begin{figure}[h] 
\includegraphics[width=0.90\textwidth]{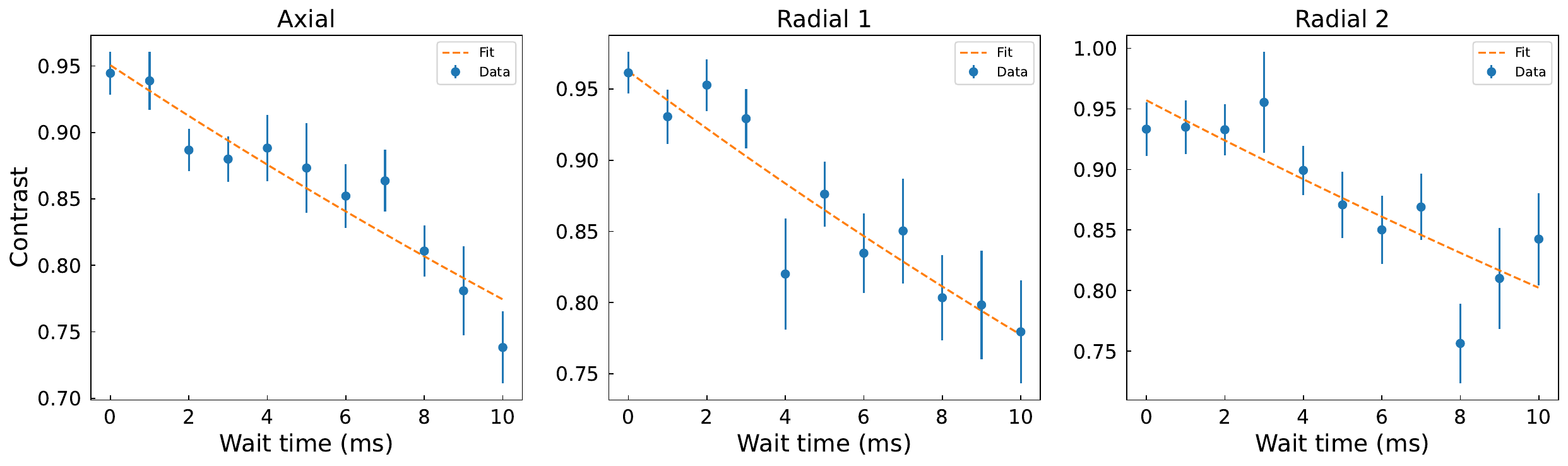}
\caption{Contrast measurements against wait times for each of the three motional modes of a single trapped ion. Each point is calculated from a 36 point scan of the second carrier phase, with each point on the phase scan equivalent to 100 experimental repetitions. Error bars are generated from sinusoidal fit errors for each of the carrier phase scans. We fit the data points to an exponentially decaying function and obtain the estimated $1/e$ coherence times for our three motional modes as $\{49(7), 47(7), 60(20)\}$ \si{ms} for the modes $\{Ax, R1, R2\}$ respectively from the fit.} 
\label{fig:coherence contrast}
\end{figure}

As shown in Figure \ref{fig:coherence contrast}, the available measurement range is insufficient to reliably determine the motional coherence times from exponential fits to the contrast decay. Nevertheless, the measured contrast is above 0.70 for all three modes at 10 ms, indicating that the motional coherence times exceed the experimentally probed timescale of 10 ms. This is sufficient for our purposes as our experimental sequence duration, which includes the carrier qubit rotations, CPG protocol and characteristic function measurements, is below 1 ms. \\

We measure the motional heating rates by driving a $\pi$ pulse on both the RSB and BSB and taking the ratio of these two excitation probabilities \cite{leibfried2003}. This ratio is used to calculate the average excitation number $(\bar{n})$. We introduce a variable wait time before the sideband pulses and measure $\bar{n}$ as a function of wait time. A linear fit to $\bar{n}$ yields heating rates of $\{7.0(4),3.0(2),5.8(3)\}$ phonons per second for the $\{Ax,R1,R2\}$ motional modes, respectively. \\

\begin{figure}[h] \label{fig:heating rate}
\includegraphics[width=0.90\textwidth]{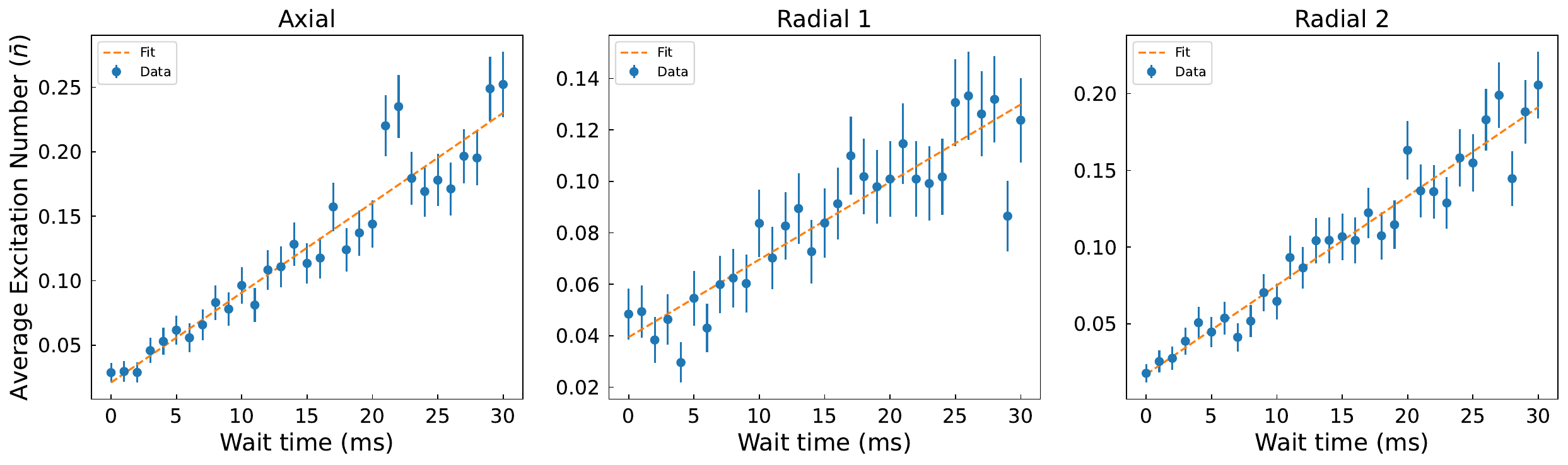}
\caption{Average excitation number $(\bar{n})$ against wait times for each of the three motional modes of a single trapped ion. Each point is calculated from 1000 experimental repetitions of the BSB drive and another 1000 experimental repetitions of the RSB drive. Error bars are generated from errors inherent in qubit spin measurements. We fit the data points to a linear function.} 
\end{figure}

\newpage

We estimate our qubit dark-state measurement error by optically pumping our qubit state to the $\ket{g} \equiv \ket{F=0,m_{F}=0}$ state. We then measure this state for 10600 experimental repetitions to obtain an error of $1.3(1)\%$. To estimate the bright-state measurement error, we first optically pump the qubit into $\ket{g}$ and then drive a microwave transition from the $\ket{g} \equiv \ket{F=0,m_{F}=0}$ state to the $\ket{e} \equiv \ket{F=1,m_{F}=0}$ state. We then measure this state for 12000 experimental repetitions to obtain an error of $2.0(1)\%$. 

\section{Simulation and experimental results for $N=6$ and $N=10$ CPG protocol}

\begin{figure}[h] \label{fig:6 combined}
\includegraphics[width=0.99\textwidth]{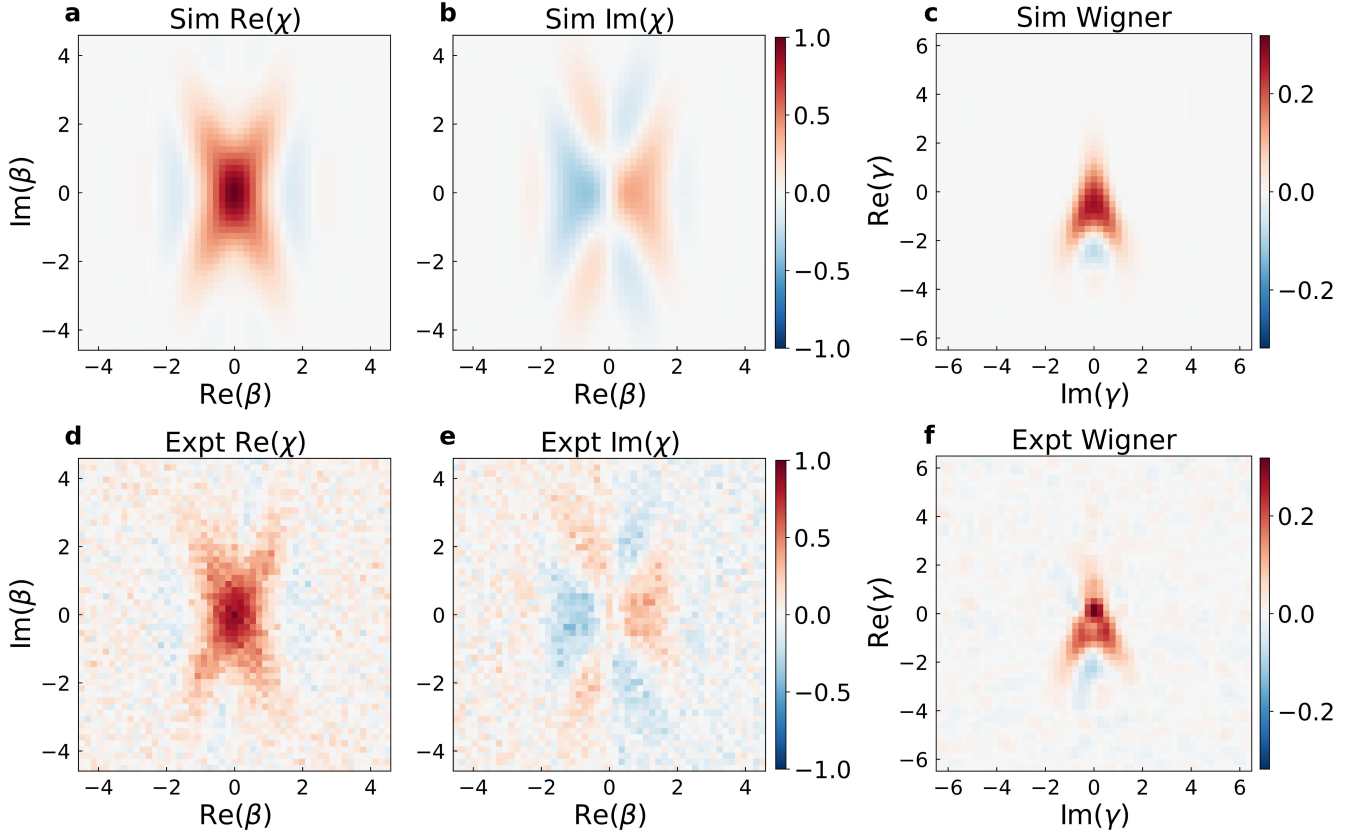}
\caption{Simulated (a) real and (b) imaginary components of the characteristic function along with the (c) simulated Wigner function describing our output state after the $N=6$ CPG protocol. Experimentally measured (d) real and (e) imaginary components of the characteristic function along with the reconstructed (f) Wigner function describing our output state after the $N=6$ CPG protocol. We calculate the state overlap between the experimental Wigner function and the simulated Wigner function in (c) to be 0.899(2).} 
\end{figure}

\begin{figure}[h] \label{fig:10 combined}
\includegraphics[width=0.99\textwidth]{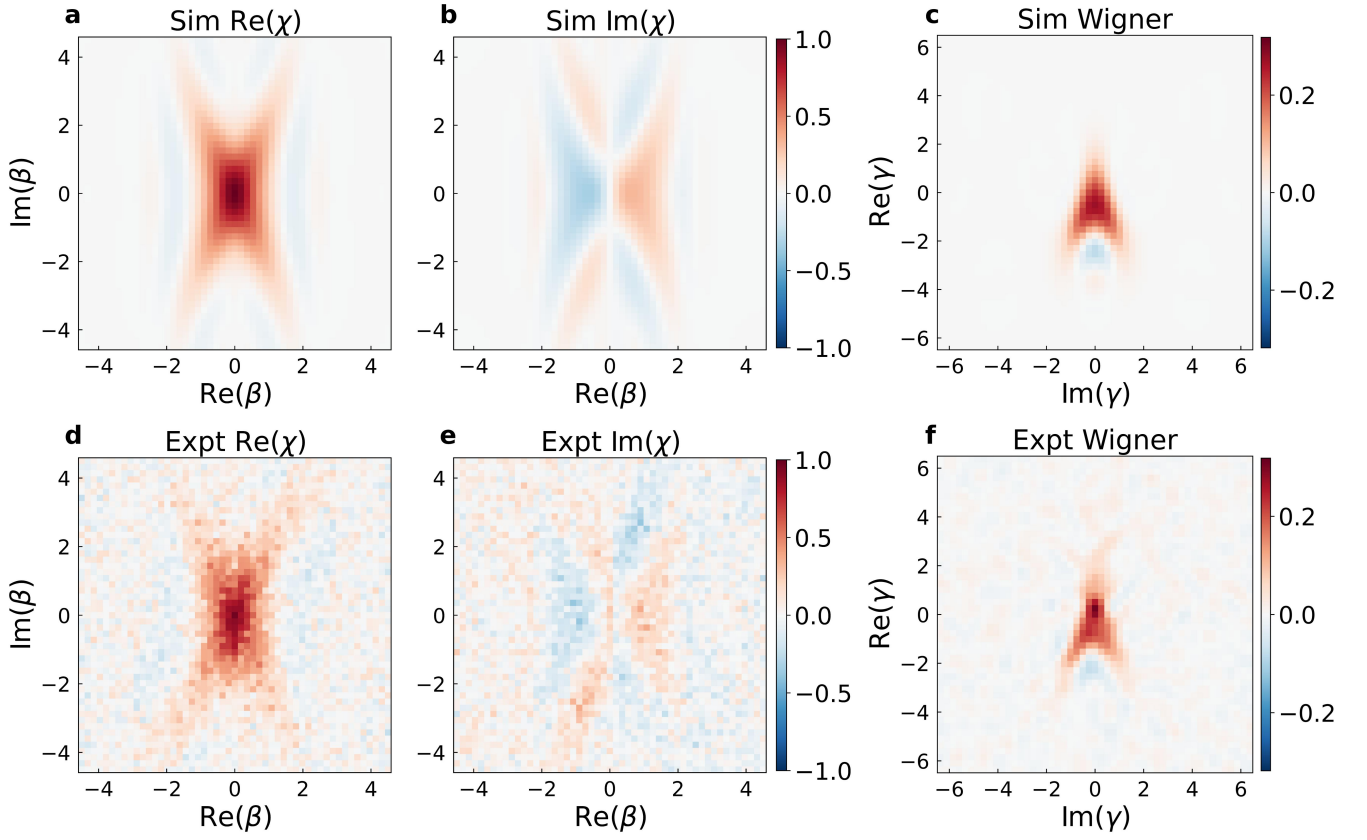}
\caption{Simulated (a) real and (b) imaginary components of the characteristic function along with the (c) simulated Wigner function describing our output state after the $N=10$ CPG protocol. Experimentally measured (d) real and (e) imaginary components of the characteristic function along with the reconstructed (f) Wigner function describing our output state after the $N=10$ CPG protocol. We calculate the state overlap between the experimental Wigner function and the simulated Wigner function in (c) to be 0.868(2).} 
\end{figure}

\newpage
\clearpage
\section{Simulation and experimental results for non-vacuum input states}

\begin{figure}[h] 
\includegraphics[width=0.99\textwidth]{right_combined.pdf}
\caption{Exact unitary simulation of the (a) real and (b) imaginary components of the characteristic function along with the exact unitary simulation of the (c) Wigner function describing our input state of a coherent state of $\alpha=0.50$ undergoing evolution under the exact cubic unitary. Simulated (d) real and (e) imaginary components of the characteristic function along with the (f) simulated Wigner function describing our output state after the $N=14$ CPG protocol with the same input state. Experimentally measured (g) real and (h) imaginary components of the characteristic function along with the reconstructed (i) Wigner function describing our output state after the $N=14$ CPG protocol with the same input state. We calculate the state overlap between the experimental Wigner function and the simulated Wigner function in (f) to be 0.928(2).} 
\label{fig:right combined}
\end{figure}

\begin{figure}[t] 
\includegraphics[width=0.99\textwidth]{left_combined.pdf}
\caption{Exact unitary simulation of the (a) real and (b) imaginary components of the characteristic function along with the exact unitary simulation of the (c) Wigner function describing our input state of a coherent state of $\alpha=-0.50$ undergoing evolution under the exact cubic unitary. Simulated (d) real and (e) imaginary components of the characteristic function along with the (f) simulated Wigner function describing our output state after the $N=14$ CPG protocol with the same input state. Experimentally measured (g) real and (h) imaginary components of the characteristic function along with the reconstructed (i) Wigner function describing our output state after the $N=14$ CPG protocol with the same input state. We calculate the state overlap between the experimental Wigner function and the simulated Wigner function in (f) to be 0.871(2).} 
\label{fig:left combined}
\end{figure}

\begin{figure}[t] 
\includegraphics[width=0.99\textwidth]{down_combined.pdf}
\caption{Exact unitary simulation of the (a) real and (b) imaginary components of the characteristic function along with the exact unitary simulation of the (c) Wigner function describing our input state of a coherent state of $\alpha=-0.50i$ undergoing evolution under the exact cubic unitary. Simulated (d) real and (e) imaginary components of the characteristic function along with the (f) simulated Wigner function describing our output state after the $N=14$ CPG protocol with the same input state. Experimentally measured (g) real and (h) imaginary components of the characteristic function along with the reconstructed (i) Wigner function describing our output state after the $N=14$ CPG protocol with the same input state. We calculate the state overlap between the experimental Wigner function and the simulated Wigner function in (f) to be 0.894(2).} 
\label{fig:down combined}
\end{figure}

\begin{figure}[t] 
\includegraphics[width=0.99\textwidth]{up_combined.pdf}
\caption{Exact unitary simulation of the (a) real and (b) imaginary components of the characteristic function along with the exact unitary simulation of the (c) Wigner function describing our input state of a coherent state of $\alpha=0.50i$ undergoing evolution under the exact cubic unitary. Simulated (d) real and (e) imaginary components of the characteristic function along with the (f) simulated Wigner function describing our output state after the $N=14$ CPG protocol with the same input state. Experimentally measured (g) real and (h) imaginary components of the characteristic function along with the reconstructed (i) Wigner function describing our output state after the $N=14$ CPG protocol with the same input state. We calculate the state overlap between the experimental Wigner function and the simulated Wigner function in (f) to be 0.931(2).} 
\label{fig:up combined}
\end{figure}

\clearpage

\section{Derivation of Non-linear squeezing for cubic case using characteristic function} \label{sec: chr func NLS}
Define the position and momentum operators:

\begin{align*}
\begin{array}{cc}
    \hat{a} &= \frac{g}{2}\qty(\hat{x}+i\hat{p})\\
    \hat{a}^\dagger &= \frac{g}{2}\qty(\hat{x}-i\hat{p})
\end{array}
\quad \implies \quad
\begin{array}{cc}
    \hat{x} &= \frac{1}{g}\qty(\hat{a}+\hat{a}^\dagger)\\
    \hat{p} &= \frac{1}{ig}\qty(\hat{a}-\hat{a}^\dagger),
\end{array}
\end{align*}
where $g$ is a scaling factor. When $g = \sqrt{2}$, we obtain the canonical relation between the ladder operators and the position and momentum operators.\\

Next, define $\beta$ in terms of its real and imaginary parts:
\begin{align*}
    \beta = b_1+ib_2, \quad \{b_1,b_2\}\in \mathbb{R}.
\end{align*}
Then, the characteristic function can be re-written in the position and momentum operator as such:
\begin{align*}
    \chi(b_1,b_2) &= \expval{\exp((b_1+ib_2) \hat{a}^\dagger-(b_1-ib_2))\hat{a})}_\rho\\
    &=\expval{\exp( b_1(\hat{a}^\dagger-\hat{a})+ib_2(\hat{a}^\dagger+\hat{a}))}_\rho\\
    &=\expval{\exp( ig(b_2\hat{x}-b_1\hat{p}))}_\rho.
\end{align*}
Re-writing $V^3[\rho](\lambda)$ to its specific components
\begin{align*}
    V^3[\rho](\lambda) &= Var(\hat{p}_{NLQ})\\
    &= \expval{(\hat{p}-3\lambda \hat{x}^2)^2}_\rho-\expval{\hat{p}-3\lambda \hat{x}^2}^2_\rho\\
    &=\expval{\hat{p}^2}_\rho-3\lambda \expval{\hat{p}\hat{x}^2+\hat{x}^2\hat{p}}_\rho+9\lambda^2\expval{\hat{x}^4}_\rho-\qty(\expval{\hat{p}}_\rho-3\lambda\expval{\hat{x}^2}_\rho)^2.
\end{align*}
Using the property of characteristic function, we can obtain the components as such \cite{gerry_introductory_2005}:
\begin{align} 
    \expval{\hat{p}}_\rho &= -\frac{1}{ig}\frac{\partial\chi(b_1,b_2)}{\partial b_1}\bigg\vert_{b_1,b_2 = 0} \label{eq:derivatives1} \\
    \expval{\hat{p}^2}_\rho &= -\frac{1}{g^2}\frac{\partial^2\chi(b_1,b_2)}{\partial b_1^2}\bigg\vert_{b_1,b_2 = 0} \label{eq:derivatives2} \\
    \expval{\hat{x}^2}_\rho &= -\frac{1}{g^2}\frac{\partial^2\chi(b_1,b_2)}{\partial b_2^2}\bigg\vert_{b_1,b_2 = 0} \label{eq:derivatives3} \\
    \expval{\hat{x}^4}_\rho &= \frac{1}{g^4}\frac{\partial^4\chi(b_1,b_2)}{\partial b_2^4}\bigg\vert_{b_1,b_2 = 0} \label{eq:derivatives4} \\
    \expval{\hat{p}\hat{x}^2+\hat{x}^2\hat{p}}_\rho&=\frac{2}{ig^3}\frac{\partial^3\chi(b_1,b_2)}{\partial b_1\partial b_2^2}\bigg\vert_{b_1,b_2 = 0} . \label{eq:derivatives5} 
\end{align}
Due to the existence of cross terms between $\hat{x}$ and $\hat{p}$, additional proof is required for the last relation (\ref{eq:derivatives5}).
\begin{align}
    \frac{\partial^3\chi(b_1,b_2)}{\partial b_1\partial b_2^2}\bigg\vert_{a,b = 0} &= \frac{ig^3}{3}\expval{\hat{p}\hat{x}^2+\hat{x}^2\hat{p}+\hat{x}\hat{p}\hat{x}}_\rho \nonumber\\
    &=\frac{ig^3}{3}\expval{\hat{p}\hat{x}^2+\hat{x}^2\hat{p}+\frac{1}{2}\qty(\hat{p}\hat{x}^2+i\hat{x}+\hat{x}^2\hat{p}-i\hat{x})}_\rho \nonumber \\
    &=\frac{ig^3}{2}\expval{\hat{p}\hat{x}^2+\hat{x}^2\hat{p}}_\rho ,\label{eq:derivatives6} 
\end{align}
where the first line comes about due to the fact that $ \frac{\partial^3\chi(b_1,b_2)}{\partial b_1\partial b_2^2}\vert_{b_1,b_2 = 0} $ has 3 possible permutations of $\hat{p}$ and $\hat{x}$, thus the need to account for all of them and taking their average. This is also known as Weyl ordering. In the 2nd line, the commutator $\qty[\hat{x}, \hat{p}] = i$ was used.

\section{Derivation of the fitting function for $\chi(b_1,b_2)$} \label{fitting function}
To begin, we note that
\begin{align*}
    \chi(b_1,b_2) &= \expval{\exp( ig(b_2\hat{x}-b_1\hat{p}))}_\rho\\
    &=\expval{\exp(-i\kappa \hat{x}^3)\exp( ig(b_2\hat{x}-b_1\hat{p}))\exp(i\kappa \hat{x}^3)}_{vac}.
\end{align*}
It can be seen that the characteristic function $\chi$ can be found by finding the vacuum expectation value of the operator $\exp(-i\kappa \hat{x}^3)\exp( ig(b_2\hat{x}-b_1\hat{p}))\exp(i\kappa \hat{x}^3)$. Hence, the problem reduces to determining how the operator acts on the vacuum state.\\

One straightforward way to do so is to disentangle the operator and convert into its ladder operator form. Before starting the disentangling process, note that by applying the Baker–Campbell–Hausdorff (BCH) formula once, the cubic term is reduced to a quadratic term, which would slightly simplify the disentangling process. 
\begin{align*}
    \exp(-i\kappa \hat{x}^3)\exp( ig(b_2\hat{x}-b_1\hat{p}))\exp(i\kappa \hat{x}^3)&=\exp(\exp(-i\kappa \hat{x}^3) ig(b_2\hat{x}-b_1\hat{p})\exp(i\kappa \hat{x}^3))\\
    &=\exp(ig(b_2\hat{x}-b_1\hat{p})-3ib_1g\kappa \hat{x}^2)\\
    &=\exp(ig(b_2\hat{x}-b_1\hat{p}-3b_1\kappa \hat{x}^2)),
\end{align*}
since, 
\begin{align*}
    \qty[-i\kappa\hat{x}^3,ig(b_2\hat{x}-b_1\hat{p}))]&=-b_1g\kappa\qty[\hat{x}^3,\hat{p}]\\
    &=-3ib_1g\kappa \hat{x}^2,
\end{align*}
which also implies that $\qty[-i\kappa\hat{x}^3,\qty[-i\kappa\hat{x}^3,ig(b_2\hat{x}-b_1\hat{p}))]]=0$. \\

Now, we begin the disentangling process. Applying Zassenhaus formula, 
\begin{align*}
    \exp(ig(b_2\hat{x}-b_1\hat{p}-3b_1\kappa \hat{x}^2))&=\exp(ig(b_2\hat{x}-b_1\hat{p}))\exp(-3igb_1\kappa\hat{x}^2)\exp(-3i\kappa g^2b_1^2\hat{x})\exp(-i\kappa g^3b_1^3),
\end{align*}
where the following commutators are applied, 
\begin{align*}
    \qty[b_2\hat{x}-b_1\hat{p}, -3b_1\kappa\hat{x}^2] &= -6i\kappa b_1^2 \hat{x}\\
    \qty[b_2\hat{x}-b_1\hat{p}, -6i\kappa b_1^2 \hat{x}] &= 6\kappa b_1^3.
\end{align*}
To clean things up a little, label $\beta = b_1+ib_2$, $\epsilon = -3i\kappa gb_1^2$ and $A = -3i\kappa b_1/g$,
\begin{align*}
    \exp(ig(b_2\hat{x}-b_1\hat{p}-3b_1\kappa \hat{x}^2)) &= \exp(-i\kappa g^3b_1^3)\exp(\beta \hat{a}^\dagger - \beta^*\hat{a})\exp(\epsilon\hat{a}^\dagger-\epsilon^*\hat{a})\exp(A(\hat{a}^\dagger{}^2+\hat{a}^2+2\hat{a}^\dagger\hat{a}+1))\\
    &=\exp(-i\kappa g^3b_1^3)D(\beta)D(\epsilon)\exp(A(\hat{a}^\dagger{}^2+\hat{a}^2+2\hat{a}^\dagger\hat{a}+1))\\
    &=\exp(-i\kappa g^3b_1^3)\exp(3i\kappa g b_1^3)D(\beta+\epsilon)\exp(A(\hat{a}^\dagger{}^2+\hat{a}^2+2\hat{a}^\dagger\hat{a}+1))\\
    &=\exp(i\kappa g (3-g^2)b_1^3)D(\beta+\epsilon)\exp(A(\hat{a}^\dagger{}^2+\hat{a}^2+2\hat{a}^\dagger\hat{a}+1)).\\
\end{align*}
The disentangling of $D(\beta+\epsilon)$ is straightforward,
\begin{align*}
    D(\beta+\epsilon) &= \exp(-\frac{1}{2}\qty|\beta+\epsilon|^2)\exp((\beta+\epsilon)\hat{a}^\dagger)\exp(-(\beta+\epsilon)^*\hat{a}).
\end{align*}
Now, we shall determine the disentanglement of $\exp(A(\hat{a}^\dagger{}^2+\hat{a}^2+2\hat{a}^\dagger\hat{a}+1))$. Let,
\begin{align*}
    F(s) = \exp(As(\hat{a}^\dagger{}^2+\hat{a}^2+2\hat{a}^\dagger\hat{a}+1)).
\end{align*}
We choose the ansatz,
\begin{align*}
    F(s) &= \exp(B(s)\hat{a}^\dagger{}^2)\exp(C(s)(2\hat{a}^\dagger\hat{a}+1))\exp(D(s)\hat{a}^2),\hspace{5mm}\text{with } B(0) = C(0) = D(0) = 0.
\end{align*}
Differentiating both $F(s)$ w.r.t $s$, we require
\begin{align*}
    A(\hat{a}^\dagger{}^2+\hat{a}^2+2\hat{a}^\dagger\hat{a}+1) &= \dot{B}(s)\hat{a}^\dagger{}^2+\dot{C}(s)e^{B(s)\hat{a}^\dagger{}^2}(2\hat{a}^\dagger\hat{a}+1)e^{-B(s)\hat{a}^\dagger{}^2}+\dot{D}(s)e^{B(s)\hat{a}^\dagger{}^2}e^{C(s)(2\hat{a}^\dagger\hat{a}+1)}\hat{a}^2e^{-C(s)(2\hat{a}^\dagger\hat{a}+1)}e^{-B(s)\hat{a}^\dagger{}^2}\\
    &=\dot{B}(s)\hat{a}^\dagger{}^2+\dot{s}(s)\qty(2\hat{a}^\dagger\hat{a}+1-4B(s)\hat{a}^\dagger{}^2)+e^{-4C(s)}\dot{D}(s)\qty(\hat{a}^2-2B(s)(2\hat{a}^\dagger\hat{a}+1)+4B^2(s)\hat{a}^\dagger{}^2).
\end{align*}
By comparing coefficients, we have 
\begin{align*}
    \dot{B}(s)-4B(s)\dot{C}(s)+4e^{-4C(s)}B^2(s)\dot{D}(s) &= A\\
    \dot{C}(s)-2e^{-4C(s)}B(s)\dot{D}(s) &= A\\
    e^{-4C(s)}\dot{D}(s)&=A.
\end{align*}
Solving them gives
\begin{align*}
     B(s) &= \frac{As}{1-2As}\\
     C(s) &=-\frac{1}{2}\ln|1-2As|\\
     D(s) &= \frac{As}{1-2As}.
\end{align*}
Thus, setting $s=1$, 
\begin{align*}
    F(1) &=\exp(A(\hat{a}^\dagger{}^2+\hat{a}^2+2\hat{a}^\dagger\hat{a}+1))\\
    &=\exp(\frac{A}{1-2A}\hat{a}^\dagger{}^2)\exp(-\frac{1}{2}\ln|1-2A|(2\hat{a}^\dagger\hat{a}+1))\exp(\frac{A}{1-2A}\hat{a}^2)\\
    &=\frac{1}{\sqrt{|1-2A|}}\exp(\frac{A}{1-2A}\hat{a}^\dagger{}^2)\exp(-\ln|1-2A|\hat{a}^\dagger\hat{a})\exp(\frac{A}{1-2A}\hat{a}^2).
\end{align*}
Putting everything back together, 
\begin{align*}
    \expval{\exp(ig(b_2\hat{x}-b_1\hat{p}-3b_1\kappa \hat{x}^2))}_{vac} &= \frac{\exp(i\kappa g (3-g^2)b_1^3)\exp(-\frac{1}{2}\qty|\beta+\epsilon|^2)}{\sqrt{|1-2A|}}\bigg\langle\exp((\beta+\epsilon)\hat{a}^\dagger)\exp(-(\beta+\epsilon)^*\hat{a})\\
    &\phantom{\frac{\exp(-\frac{1}{2}\qty|\beta+\epsilon|^2)}{\sqrt{|1-2A|}}}\times\exp(\frac{A}{1-2A}\hat{a}^\dagger{}^2)\exp(-\ln|1-2A|\hat{a}^\dagger\hat{a})\exp(\frac{A}{1-2A}\hat{a}^2)\bigg\rangle_{vac}\\
    &=\frac{\exp(i\kappa g (3-g^2)b_1^3)\exp(-\frac{1}{2}\qty|\beta+\epsilon|^2)}{\sqrt{|1-2A|}}\expval{\exp(-(\beta+\epsilon)^*\hat{a})\exp(\frac{A}{1-2A}\hat{a}^\dagger{}^2)}_{vac}\\
    &=\frac{\exp(i\kappa g (3-g^2)b_1^3)\exp(-\frac{1}{2}\qty|\beta+\epsilon|^2)}{\sqrt{|1-2A|}}\\
    &\phantom{\frac{\exp(-\frac{1}{2}\qty|\beta+\epsilon|^2)}{\sqrt{|1-2A|}}}\times\sum_{m=0}^\infty\sum_{n=0}^\infty\frac{1}{n!}\frac{1}{m!}\qty(\frac{A}{1-2A})^n\qty(-(\beta+\epsilon)^*)^m\sqrt{(2n)!}\sqrt{\frac{(2n)!}{(2n-m)!}}\delta_{2n,m}\\
    &=\frac{\exp(i\kappa g (3-g^2)b_1^3)\exp(-\frac{1}{2}\qty|\beta+\epsilon|^2)}{\sqrt{|1-2A|}}\sum_{n=0}^\infty\frac{1}{n!}\qty(((\beta+\epsilon)^*)^2\frac{A}{1-2A})^n\\
    &=\frac{1}{\sqrt{|1-2A|}}\exp(i\kappa g (3-g^2)b_1^3)\exp(-\frac{1}{2}\qty|\beta+\epsilon|^2)\exp(\frac{A((\beta+\epsilon)^*)^2}{1-2A})\\
    &=\frac{1}{\sqrt{|1-2A|}}\exp(i\kappa g (3-g^2)b_1^3)\exp(-\frac{1}{2}\frac{\qty|\beta+\epsilon|^2}{1-2A})\exp(\frac{A(\beta+\epsilon)^*((\beta+\epsilon)^*+(\beta+\epsilon))}{1-2A})\\
    &=\frac{1}{\sqrt{|1-2A|}}\exp(i\kappa g (3-g^2)b_1^3)\exp(-\frac{1}{2}\frac{\qty|\beta+\epsilon|^2}{1-2A})\exp(\frac{2b_1A(\beta+\epsilon)^*}{1-2A})\\
    &=\frac{1}{\sqrt{|1-2A|}}\exp(i\kappa g (3-g^2)b_1^3)\exp(-\frac{1}{2}\frac{\qty(\beta+\epsilon)^*(\beta+\epsilon-4b_1A)}{1-2A})\\
    &=\frac{1}{\sqrt{|1-2A|}}\exp(i\kappa g (3-g^2)b_1^3)\\
    &\hspace{3cm}\times\exp(-\frac{1}{2}\frac{b_1^2+b_2^2-4b_1^2A-(2g^2-4)b_1b_2Ai-(g^2-4)g^2b_1^2A^2}{1-2A}).
\end{align*}
Rewrite the function in the parameters of only $b_1$ and $b_2$ and relabel all other constants into fitting parameters $c_i$, 
\begin{align*}
\expval{\exp(ig(b_2\hat{x}-b_1\hat{p}-3b_1\kappa \hat{x}^2))}_{vac} &= \frac{1}{(c_1+c_2b_1^2)^{1/4}}\exp(\frac{c_3b_1^2+c_4b_2^2+c_5b_1^2b_2+c_6b_1^4}{c_1+c_2b_1^2})\\
&\phantom{==}\times\exp(i\qty(\frac{c_{7}b_1b_2^2+c_{8}b_1^3+c_{9}b_1^3b_2+c_{10}b_1^5}{c_1+c_2b_1^2}+c_{11}\tan^{-1}\qty(c_{12}b_1))).
\end{align*}
Let 
\begin{align*}
    \Xi_R &= \frac{c_3b_1^2+c_4b_2^2+c_5b_1^2b_2+c_6b_1^4}{c_1+c_2b_1^2}\\
    \Xi_I &= \frac{c_{7}b_1b_2^2+c_{8}b_1^3+c_{9}b_1^3b_2+c_{10}b_1^5}{c_1+c_2b_1^2}+c_{11}\tan^{-1}\qty(c_{12}b_1).
\end{align*}
The real and imaginary parts are therefore
\begin{align}
    \Re(\expval{\exp(ig(b_2\hat{x}-b_1\hat{p}-3b_1\kappa \hat{x}^2))}_{vac}) &= \frac{1}{(c_1+c_2b_1^2)^{1/4}}\exp(\Xi_R)\cos(\Xi_I)\\
    \Im(\expval{\exp(ig(b_2\hat{x}-b_1\hat{p}-3b_1\kappa \hat{x}^2))}_{vac})&=\frac{1}{(c_1+c_2b_1^2)^{1/4}}\exp(\Xi_R)\sin(\Xi_I)
\end{align}

\clearpage
\section{Non-linear Squeezing (NLS)}

We calculate and analyze NLS based on the methods laid out in \cite{Moore_2019}. However, rather than calculating NLS from Wigner functions, we calculate NLS directly from our experimental characteristic functions via the expectation values derived in Section \ref{sec: chr func NLS}. In order to obtain these expectation values, we calculated the higher-order derivatives of the characteristic functions in \cref{eq:derivatives1,eq:derivatives2,eq:derivatives3,eq:derivatives4,eq:derivatives5}. \\

To evaluate these derivatives, we fit the experimental characteristic function to a two-dimensional fitting function derived in Section \ref{fitting function}. For the derivatives in \cref{eq:derivatives1,eq:derivatives2,eq:derivatives3,eq:derivatives4}, we plot a one-dimensional curve from the fitting function using 10001 points, compared to the 51 points in the experimental data, over the same range $\operatorname{Re}[\beta],\operatorname{Im}[\beta]\in[-4.5,4.5]$. The required derivatives are then evaluated numerically from these higher-resolution fit curves. For the cross-term derivative (\ref{eq:derivatives5}), we evaluate the fitted function on a $101\times101$ grid over the range $\operatorname{Re}[\beta],\operatorname{Im}[\beta]\in[-0.045,0.045]$. The resulting two-dimensional fitted curve is differentiated numerically once with respect to $\operatorname{Re}[\beta]$ and twice with respect to $\operatorname{Im}[\beta]$ to obtain the derivative in (\ref{eq:derivatives5}). \\

\begin{figure}[h] 
\includegraphics[width=0.50\textwidth]{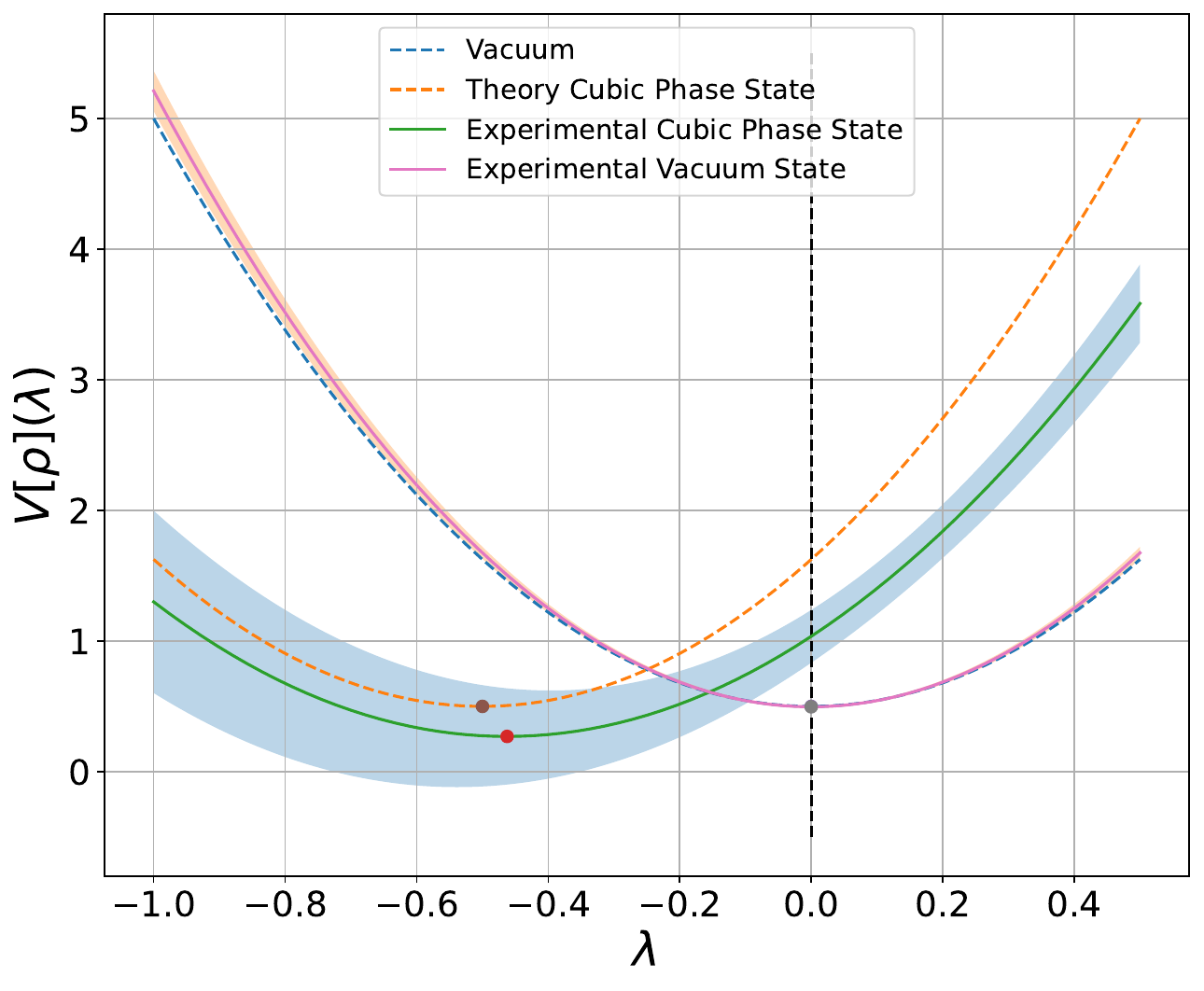}
\caption{NLS of our experimental characteristic function output from the $N=14$ CPG protocol as reported previously in the main text. Here we report the comparison between the theoretical cubic phase state NLS (orange dashed line) and the experimental NLS (green solid line) with its experimental uncertainty (blue shaded area). We also compare the NLS between the theoretical vacuum state (blue dashed line) and the experimental vacuum state (magenta solid line) with its experimental uncertainty (orange shaded area). The markers indicate the minimum points for the theoretical cubic phase state NLS (brown), experimental cubic phase state NLS (red). Both vacuum state NLS curves have very similar minimum points as such share a marker (grey). } 
\label{fig:cubicity detailed}
\end{figure}

We estimate the experimental uncertainty of NLS via a Monte Carlo method. For each iteration cycle, we inject noise into the experimental characteristic function which follows the statistics from the binomial distribution where the number of trials is set to 200 and the mean is set to the experimentally measured average excitation probability of the qubit. The resulting ``noisy" characteristic function is refitted to the same fitting function, from which a new set of expectation values and the corresponding NLS curve are obtained. We apply the same Monte Carlo procedure to the experimentally measured vacuum state but instead of the fitting functions in Section \ref{fitting function}, we use Gaussian fitting functions for $\operatorname{Re}[\chi]$ and a constant offset function for $\operatorname{Im}[\chi]$. Each Monte Carlo iteration therefore produces a different NLS curve. We perform 1066 iterations for the experimental cubic phase state and 1100 iterations for the experimental vacuum state to obtain the average NLS curve and its 1 standard deviation uncertainty as shown in Figure \ref{fig:cubicity detailed}. \\ 

To determine the value for the minimum values, we take the minimum points for all generated curves then report the average NLS and $\lambda$ values along with their associated 1 standard deviation uncertainty. To obtain the shaded uncertainty regions in Figure \ref{fig:cubicity detailed}, we calculate the average NLS values $(\mu_{NLS})$ along with its associated 1 standard deviation uncertainty $(\sigma_{NLS})$ for every value of $\lambda$. The upper and lower bounds of the shaded region are then given by $\mu_{{NLS}}+\sigma_{{NLS}}$ and $\mu_{{NLS}}-\sigma_{{NLS}}$, respectively. \\

For a state to exhibit nonlinear squeezing, the NLS curve for the cubic phase state must be suppressed below that of the NLS of the vacuum state for some range of $\lambda$ \cite{Moore_2019}. As shown in Figure \ref{fig:cubicity detailed} and as reported in the main text, the theoretical vacuum state curve intersects the upper limit of the shaded area at $\lambda=-0.227(2)$, where the quoted uncertainty includes the resolution error. We also calculate the intersection of the lower limit of the experimental vacuum state shaded area with the upper limit of the experimental cubic phase state shaded area to be $\lambda=-0.228(2)$. This shows good agreement between the theoretical and experimental vacuum states. \\

The experimental vacuum state NLS curve reaches a minimum of $0.497(9)$ at $\lambda\approx0$, in close agreement with the theoretical minimum of $0.5$ at $\lambda=0$. The experimental cubic phase state reaches a minimum NLS of $0.27(39)$ at $\lambda=-0.46(8)$. The location of this minimum is consistent with the theoretical prediction at $\lambda=-0.5$, as reported in the main text. At this minimum point of $\lambda=-0.46(8)$, the NLS of the experimental cubic phase state is suppressed below the theoretical vacuum state by $3.2\sigma_{NLS}$. \\

The theoretical NLS curves for the vacuum and cubic phase states shown in Figure \ref{fig:cubicity detailed} follow equation (\ref{eq:theoretical cubicity}) \cite{Moore_2019} where $\kappa = -0.5$ for the cubic phase state chosen for this project and $\kappa=0$ for vacuum, \\

\begin{equation} \label{eq:theoretical cubicity}
    V\left[\ket{\kappa}\bra{\kappa}\right](\lambda)=\frac{1}{2} \left(1+9(\kappa-\lambda)^2 \right).
\end{equation}

\clearpage

\bibliographystyle{apsrev4-2}
\bibliography{apssamp.bib}

\end{document}